\documentclass[aps,amsmath,amssymb,twocolumn,longbibliography,10pt,prx,superscriptaddress]{revtex4-2}
\usepackage{graphicx}
\usepackage[colorlinks=true,linkcolor=blue,citecolor=blue,urlcolor=blue]{hyperref}
\usepackage{color}
\usepackage{multirow}
\usepackage{tabularx}

\begin{document}

\title{Ground state of the Hubbard model with spin-dependent linear potential}
\author{Jacek Dobrzyniecki}
\email{Jacek.Dobrzyniecki@fuw.edu.pl}
\affiliation{Quantum Systems Unit, Okinawa Institute of Science and Technology Graduate University, Okinawa 904-0495, Japan}
\affiliation{Faculty of Physics, University of Warsaw, Pasteura 5, 02-093 Warsaw, Poland}
\author{Thomas Busch}
\email{thomas.busch@oist.jp}
\affiliation{Quantum Systems Unit, Okinawa Institute of Science and Technology Graduate University, Okinawa 904-0495, Japan}
\date{\today}

\begin{abstract}
We investigate the competition between attractive spin-spin interactions and spin-separating external forces in the ground state of a one-dimensional Fermi-Hubbard model. We consider a lattice with open boundary conditions, subject to a linear external potential whose gradient is opposite for the two spin components, so that each spin species sees a potential minimum at a different end of the lattice. Using density-matrix renormalization group (DMRG) simulations, we map the ground-state density distributions and the number of doubly occupied sites as a function of the potential gradient $\beta$ and interaction strength. We identify three distinct regimes separated by critical threshold gradients: (i) a small-$\beta$ regime where fermion pairing remains robust against the external potential; (ii) an intermediate-$\beta$ phase-separated regime characterized by a staircase-like decrease in the doublon number, corresponding to the successive, one-by-one breaking of bound pairs; and (iii) a large-$\beta$ regime where the two spin components are completely spatially separated. We complement the numerical results with a phenomenological model and a local-density approximation analysis, from which we derive closed-form analytical estimates for these critical threshold values. We also verify that the staircase structure persists under additional harmonic confinement. Our results are directly testable in cold-atom experiments, and demonstrate that a spin-dependent linear potential enables precise, integer-level control of the number of bound fermion pairs.
\end{abstract}

\maketitle

\section{Introduction}
\label{sec:introduction}

The Fermi--Hubbard model~\cite{10.1098/rspa.1963.0204} is a minimal lattice model of interacting fermions that plays a central role in the theory of strongly correlated electrons in crystal lattices, as well as in quantum simulations with ultracold atoms in optical lattices~\cite{Lewenstein01032007,bloch2005ultracold}. The one-dimensional case is of particular theoretical interest, as it can be solved analytically via the Bethe ansatz~\cite{essler2005,RevModPhys.85.1633}. For attractive on-site interactions ($U < 0$), the model supports bound fermion pairs whose properties interpolate between tightly bound on-site doublons at strong coupling and spatially extended Cooper-like pairs at weak coupling~\cite{essler2005}. Understanding how these pairs respond to external perturbations is a central question in the study of pairing and superfluidity in low-dimensional systems.

In realistic cold-atom implementations, such lattice systems are always subject to external confinement, and inhomogeneous external potentials can modify the system properties in nontrivial ways. In particular, inhomogeneously trapped multicomponent systems can display complex phase separation, in which different phases coexist in different regions of the lattice~\cite{PhysRevA.70.053601,PhysRevLett.97.060403,schneider2008metallic,PhysRevB.109.L041404}. Lattices with a linear potential, also known as tilted lattices, are a particularly well-studied example, with numerous experimental realizations in cold-atom systems~\cite{greiner2002quantum,simon2011quantum,PhysRevLett.111.053003,PhysRevLett.111.185301,PhysRevLett.111.185302,meinert2014observation,PhysRevLett.124.043204,kennedy2015observation}. The local-density approximation (LDA) has been widely used to describe such inhomogeneous systems by mapping the local density at each point to the equation of state of the corresponding homogeneous system~\cite{HeidrichMeisner2010,PhysRevA.84.021602,PhysRevLett.99.240403,PhysRevA.83.063632,tezuka2010ground,PhysRevB.52.2504,PhysRevB.73.165120,PhysRevA.89.023635}. However, while often successful at predicting density profiles, the LDA is inherently limited in its ability to capture phenomena that arise from the discrete or correlated nature of the many-body ground state. A powerful alternative, particularly well suited to one-dimensional systems, is the density-matrix renormalization group (DMRG) numerical technique~\cite{schollwock2005density}, which allows for accurate solutions even for very large one-dimensional systems with arbitrary external potentials.

A particularly interesting setup involves a spin-dependent external potential, such as spin-dependent harmonic trapping ~\cite{PhysRevA.81.013628,PhysRevA.85.063608,wei2015phase,PhysRevLett.97.190403,PhysRevLett.117.195302}. Fundamentally, linear potentials with spin-dependent gradients allow one to carefully study the interplay between the spin-dependent potential and spin-spin interactions~\cite{PhysRevA.96.033632}.

Despite this activity, one scenario has not been systematically explored: the interplay between spin-spin attraction that generates pairing, and a spin-dependent linear potential that pushes the two components in opposite directions, on a finite one-dimensional lattice. In this work we therefore study a one-dimensional, attractive Fermi--Hubbard model, with a linear potential whose gradient is opposite for the two spin components. The resulting spin-separating forces then compete with spin-spin attraction. Physically, such a situation can be realized, for example, by applying a magnetic-field gradient to atoms in two states with opposite magnetic moments. 

Using DMRG simulations of finite systems with open boundary conditions (\emph{i.e.}, non-periodic), we map out the ground-state properties of this system as a function of the linear gradient strength. We show that the ground state undergoes abrupt transitions at specific gradient strengths, which correspond to a successive one-by-one breaking of bound fermion pairs. These transitions manifest as staircase-like jumps in the number of doubly-occupied sites, and as sudden rearrangements of the spin-resolved density profiles. Of particular importance are the threshold gradients $\beta_{c1}, \beta_{c2}$ which correspond, respectively, to the onset and the completion of this pair-breaking sequence. We show that these threshold gradient values can be estimated with a simple toy model.

Additionally, we analyze the system via the local-density-approximation (LDA) approach, and find that it partially predicts the ground-state properties but fails to capture the stepwise descent in pair number. The LDA does, however, yield an improved analytical estimate for $\beta_{c2}$ that complements the phenomenological model in the low-filling regime.

Our results demonstrate that a spin-dependent potential allows one to precisely tune the number of paired fermions in the lattice. They also reveal the relationships among interaction strength, potential gradient, and measured density. We also verify that the staircase-like pair-breaking structure persists when a harmonic trap is added, suggesting that our conclusions are robust under experimentally realistic trapping conditions. Note that, because all the phenomena we investigate rely on the formation of finite potential minima at the chain ends, the results are intrinsically tied to the finite system size and open boundaries.

This paper is organized as follows. In Sec.~\ref{sec:model}, we describe the many-body Hamiltonian and its basic many- and single-particle properties. In Sec.~\ref{sec:exact-diagonalization}, we analyze the many-body eigenspectrum of a small system to establish the key properties of the model, in particular the relationship between the number of bound pairs and the number of doubly occupied sites in various interaction regimes, and how the pairing in the ground state changes as the external potential gradient $\beta$ is tuned. In Sec.~\ref{sec:results-from-dmrg}, we study the ground-state properties of larger systems via the DMRG technique, focusing on the doublon number and density distribution under increasing $\beta$, and we define the threshold gradients $\beta_{c1}$, $\beta_{c2}$. In Sec.~\ref{sec:pair-loosening}, we check how much the average extent of a pair changes with $\beta$ over the range $\beta_{c1} < \beta < \beta_{c2}$, and we verify that the sharp changes in the doublon number are due to pair breaking rather than pair loosening. In Sec.~\ref{sec:phenomenological-model}, we develop a simple model of pair breaking that allows one to predict the ground-state properties at any $\beta$, and we use it to derive approximate expressions for $\beta_{c1}$ and $\beta_{c2}$. In Sec.~\ref{sec:lda}, we show the system density can be predicted via the local-density approximation, and use the LDA picture to derive a second expression for $\beta_{c2}$ that is more valid at low fillings. In Sec.~\ref{sec:harmonic}, we briefly examine the ground state in the presence of additional harmonic confinement. Finally, we summarize our findings in Sec.~\ref{sec:conclusion}.

\section{The model}
\label{sec:model}

To lay the foundations, let us first define the many-body Hamiltonian, then discuss the single-particle spectrum of the linear potential, and also highlight several useful symmetries.

\subsection{The many-body Hamiltonian}
\label{sec:many-body-hamiltonian}

We consider a one-dimensional Fermi-Hubbard model with attractive on-site inter-component interactions. The lattice has open (\emph{i.e.}, non-periodic) boundary conditions, and we number the sites as $1, \ldots, L$. The many-body Hamiltonian for such a system can be written as
\begin{align}
\label{eq:many-body-hamiltonian}
\hat{H} = &-t \sum_{j = 1}^{L-1} \sum_{\sigma=\uparrow,\downarrow} \left[ \hat{c}^\dagger_{j,\sigma} \hat{c}_{j+1,\sigma} + \mathrm{H.c.} \right] \nonumber \\
&+ U \sum_{j = 1}^L \hat{n}_{j,\uparrow} \hat{n}_{j,\downarrow} + \sum_{j = 1}^L \sum_{\sigma=\uparrow,\downarrow} V_{j,\sigma} \hat{n}_{j,\sigma} ,
\end{align}
where $\hat{c}_{j,\sigma}$ annihilates a fermion with spin $\sigma$ on site $j$, and $\hat{c}^\dagger_{j,\sigma}$ is the corresponding creation operator. These operators follow ordinary fermionic anticommutation relations, $\{ \hat{c}_{i,\sigma} , \hat{c}_{j,\sigma'} \} = 0$ and $\{ \hat{c}_{i,\sigma}, \hat{c}^\dagger_{j,\sigma'} \} = \delta_{ij} \delta_{\sigma \sigma'}$, and $\hat{n}_{j,\sigma} = \hat{c}^\dagger_{j,\sigma} \hat{c}_{j,\sigma}$ is the number operator. The tunneling amplitude $t$ sets the natural energy scale; in what follows, we set $t=1$ and implicitly express all energies in units of $t$. The parameter $U$ is the interaction strength. In this work we focus on attractive interactions, $U < 0$, and include also the non-interacting limit $U = 0$ for comparison.

The linear external potential $V_{j,\sigma}$ has a gradient $\beta$ and acts with opposite signs on the two spin components. It is defined as
\begin{align}
V_{j,\uparrow} &= +\beta (j - j_0), \\
V_{j,\downarrow} &= -\beta (j - j_0),
\end{align}
with $j_0 = \frac{L+1}{2}$, so that the potential is antisymmetric about the center of the lattice. For $\beta > 0$, the $\uparrow$ component sees the potential minimum $-\beta\frac{1}{2}(L-1)$ at site $1$, while the $\downarrow$ component sees an equal minimum at site $L$. We will consider systems with fixed particle numbers $N_\uparrow$ and $N_\downarrow$ for the two spin components, allowing us to define the total population $N = N_\uparrow + N_\downarrow$ and the filling $n = N/L$.

Throughout the work we focus on balanced systems ($N_\uparrow = N_\downarrow$) in the under-half-filling ($n < 1$) regime. Interestingly, as we detail in Appendix~\ref{sec:many-body-hamiltonian-symmetries}, results for $n > 1$ can be obtained directly from the corresponding under-half-filling results by exploiting the particle--hole duality. Furthermore, we only consider $\beta \ge 0$, as results for $\beta < 0$ can be obtained by exploiting the symmetries of $\hat{H}$ that we describe in Appendix~\ref{sec:many-body-hamiltonian-symmetries}.

To find the ground state of the Hamiltonian in Eq.~\eqref{eq:many-body-hamiltonian}, we use the DMRG method~\cite{schollwock2005density} and its implementation in the TeNPy library for Python~\cite{tenpy2024}. DMRG is a well-established numerical method for finding the ground states and low-energy spectra of one-dimensional lattice models.

\subsection{The one-body spectrum}
\label{sec:one-body-spectrum}

Although we will later obtain many-body ground states numerically, it is useful to first examine the single-particle spectrum analytically. This analysis highlights useful symmetries and provides insight into how changing the gradient $\beta$ affects the density distributions. Additionally, the single-particle eigenenergies will be useful in later discussion of breaking fermion pairs into single fermions.

The one-body Hamiltonian for an atom with spin $\sigma$ can be written as
\begin{align}
    \label{eq:one-body-hamiltonian}
    \hat{h}_\sigma = &-t \sum_{j=1}^{L-1} \left( |j, \sigma \rangle \langle j+1,\sigma| + |j+1, \sigma \rangle \langle j,\sigma|\right) \\
    &+ \sum_{j=1}^L V_{j,\sigma} |j,\sigma\rangle\langle j,\sigma| = \sum_{k=1}^L \mathcal{E}_{k,\sigma} |\phi_{k,\sigma}\rangle \langle \phi_{k,\sigma}|, \nonumber 
\end{align}
where $|j,\sigma\rangle = \hat{c}^\dagger_{j,\sigma} |\mathrm{vacuum}\rangle$. The eigenorbitals $|\phi_{k,\sigma}\rangle = \sum_j \phi_{k,\sigma}(j) |j,\sigma\rangle$ are labeled as $k = 1, \ldots, L$ and have corresponding eigenenergies $\mathcal{E}_{k,\sigma}$. Because the potential satisfies $V_{j,\sigma} = V_{L+1-j,-\sigma}$, a convenient symmetry exists: The Hamiltonians $\hat{h}_\uparrow$ and $\hat{h}_\downarrow$ are related by $\hat{h}_\sigma = \hat{J}^s \hat{h}_{-\sigma}\left(\hat{J}^s\right)^\dagger$, where
\begin{equation}
    \hat{J}^s = \sum_{\sigma=\uparrow,\downarrow} \sum_{j=1}^L |L+1-j, -\sigma\rangle \langle j, \sigma|.
\end{equation}
Consequently, $\hat{h}_\uparrow$ and $\hat{h}_\downarrow$ share the same eigenenergies $\mathcal{E}_{k,\uparrow} = \mathcal{E}_{k,\downarrow}$, and their orbital wave functions are related by spatial reflection: $\phi_{k,\uparrow}(j) = \phi_{k,\downarrow}(L+1-j)$.

For $\beta = 0$, finding the eigenspectrum of $\hat{h}_\sigma$ reduces to a lattice analogue of the textbook particle-in-a-box problem, and can be done trivially via diagonalization of the hopping Hamiltonian. One obtains eigenstates with wave functions $\phi_{k,\sigma}(j) = \sqrt{\frac{2}{L+1}} \sin\left(\frac{j k \pi}{L+1}\right)$ and eigenenergies $\mathcal{E}_{k, \sigma} = -2 \cos\left(\frac{k \pi}{L+1}\right)$, corresponding to $L$ distinct quasimomenta $k \pi /(L+1)$.

For $|\beta| > 0$ on an infinite chain, the system reduces to the Wannier--Stark problem, for which the spectrum is known exactly~\cite{PhysRevB.105.184307}. The eigenenergies form a Stark ladder with a constant spacing of $\beta$ between successive levels. The corresponding eigenfunctions, labeled $\ldots,{k-1},k,{k+1},\ldots$, have identical Bessel-like density envelopes, but are spatially shifted so as to be centered on sites $\ldots,{k-1},k,{k+1},\ldots$.

For open-boundary finite-length chains, the one-body eigenfunctions for $|\beta| > 0$ are given analytically by combinations of Bessel $J$-functions~\cite{stey1973wannier}. For both spin components, the lowest-energy eigenfunctions (${k \approx 1}$) are localized near the potential minimum at one of the ends of the lattice, while the highest-energy eigenfunctions (${k \approx L}$) are localized near the potential maximum at the opposite end. The orbitals localized in the bulk (${k \approx L/2}$) are largely unaffected by the boundaries, and their energies remain close to a Stark ladder: ${\mathcal{E}_{k,\uparrow} = \mathcal{E}_{k,\downarrow} \approx \beta[k-j_0]}$. By contrast, the orbitals at the band edges, which are localized near the lattice boundaries, are significantly modified relative to the infinite-chain solution, and their energies deviate from the Wannier--Stark expression. An analytical solution for all eigenenergies is available for this case~\cite{stey1973wannier}. For a lattice with an external potential $V(j) = \beta j$ and hard-wall boundaries beyond sites $j = 1 \ldots L$, the eigenenergies $\mathcal{E}$ are given by solutions of 
\begin{equation}
\label{eq:zeros-of-lommel-polynomial}
R_{L,1-\mathcal{E}/\beta}(-2/\beta) = 0,
\end{equation}
where $R_{m,\nu}(z)$ is the Lommel polynomial
\begin{equation}
R_{m,\nu}(z) = \sum\limits^{\lfloor m/2 \rfloor}_{i=0} \frac{(-1)^{m-i} (m-i)!}{i! (m-2i)!} \frac{\Gamma(\nu + m - i) }{\Gamma(\nu + i)} \left(\frac{z}{2}\right)^{2i-m},
\end{equation}
with $\lfloor x \rfloor$ being the floor function.

There is no straightforward way to find the zeros $\mathcal{E}$ of Eq.~\eqref{eq:zeros-of-lommel-polynomial}, and they must be determined numerically. In Sec.~\ref{sec:phenomenological-model}, we will be specifically interested in the lowest eigenenergy $\mathcal{E}_{1,\sigma}$ (which governs the breakup of a fermion pair into two unpaired fermions in the lowest orbitals). To obtain an approximate analytical expression for $\mathcal{E}_{1,\sigma}$ as function of $\beta$, we adapt the known result for the continuous semi-infinite Wannier--Stark system~\cite{khonina2013calculating}, where the ground state energy is $-2 - a_1 \beta^{2/3}$ and $a_1 \approx -2.338$ is the highest zero of the Airy function. Assuming that the lowest energy in the lattice case has the same $\beta^{2/3}$ scaling, we computed the lowest $\mathcal{E}$ numerically from Eq.~\eqref{eq:zeros-of-lommel-polynomial} for a range of $\beta$ values and fitted the results to the form $-2 + \mathrm{const} \times \beta^{2/3}$. This yields the approximate expression (after restoring the constant shift $-j_0\beta$):
\begin{equation}
\label{eq:lowest-energy-approximation}
    \mathcal{E}_{1,\sigma} \approx -\beta\frac{L+1}{2} -2 + 2.270 \, \beta^{2/3},
\end{equation}
which is accurate (with a relative fit residual $\lesssim 10^{-2}$) up to $\beta \approx 0.8$.

The antisymmetry of the external potential (${ V_{j,\sigma} = -V_{L+1-j,\sigma} }$) gives rise to another useful symmetry. The single-particle Hamiltonian [Eq.~\eqref{eq:one-body-hamiltonian}] anti-commutes with the operator
\begin{equation}
\label{eq:reflection-and-chirality-symmetry-operator}
    \hat{\Gamma}^R = \sum_{\sigma=\uparrow,\downarrow} \sum_{j=1}^L |L+1-j, \sigma\rangle \langle j, \sigma|\, (-1)^{L+1-j}
\end{equation}
which combines spatial reflection with a chirality operation. This anti-commutation implies that the eigenstates come in pairs: if $|\phi\rangle$ is an eigenstate of $\hat{h}_\sigma$ with energy $\varepsilon$, then $\hat{\Gamma}^R|\phi\rangle$ is an eigenstate of $\hat{h}_\sigma$ with energy $-\varepsilon$. Thus, for example, the lowest- and highest-energy orbitals have identical probability densities, but spatially reflected, and with the highest-energy wave function having an extra $-1$ phase on every other site.

Useful symmetries of the many-body Hamiltonian are briefly discussed in Appendix~\ref{sec:many-body-hamiltonian-symmetries}.

\section{Exact diagonalization of a small system}
\label{sec:exact-diagonalization}

\subsection{Example of a small system}
\label{sec:small-system}

\begin{figure}[t]
    \centering
    \includegraphics[width=1\linewidth]{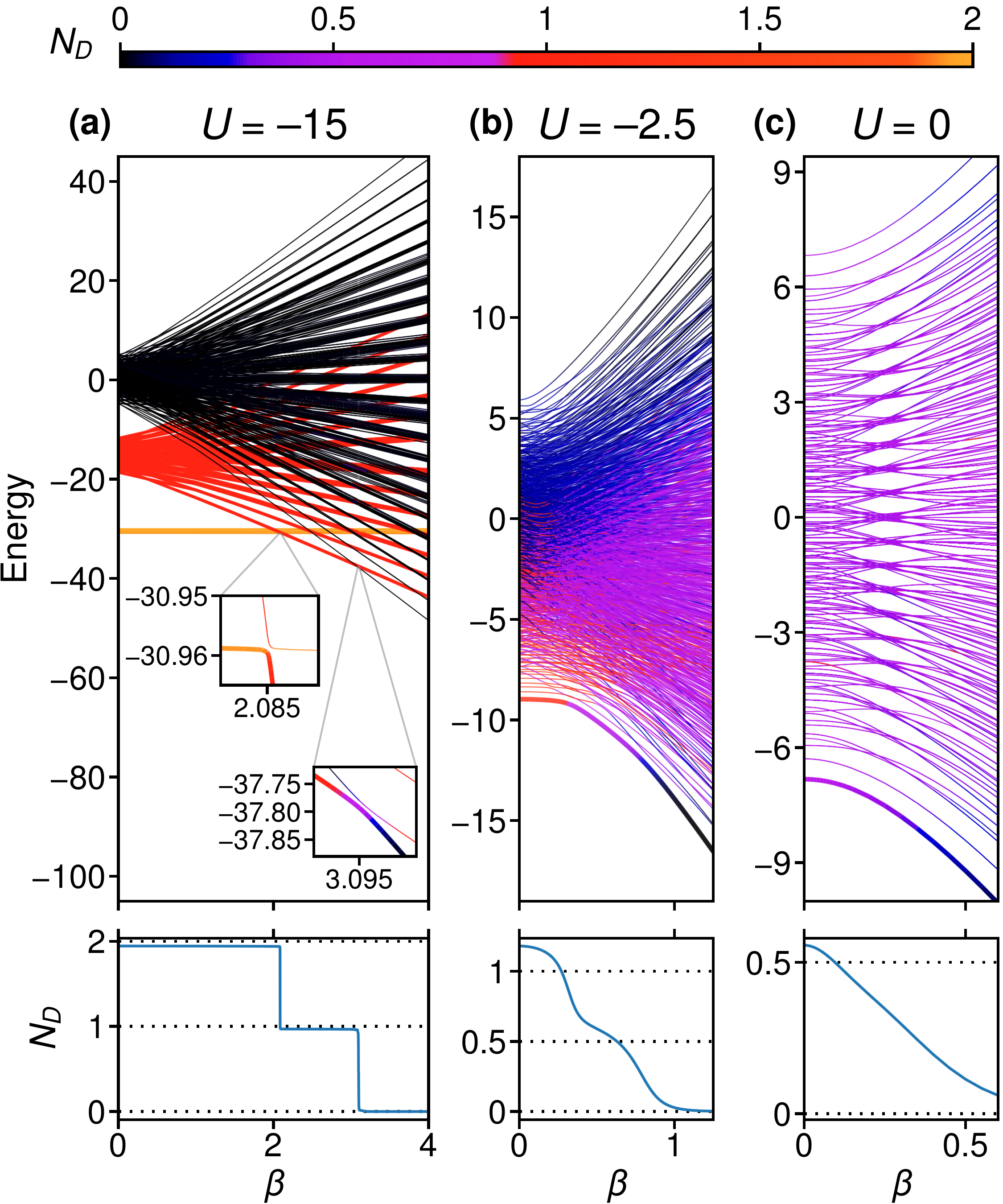}
    \caption{Top row: The many-body eigenenergy spectrum as a function of $\beta$ for (a) $U = -15$, (b) $U=-2.5$, (c) $U=0$, for a small system with size $L=8$ and population $N_\uparrow+N_\downarrow=2+2$. The colors indicate the doublon number $N_D$ in each eigenstate (as per the color scale at the top). In (a), the ground-state anticrossings with nearby eigenstates are shown closely in the insets. For clarity, the ground state is plotted with a thicker line in (b), (c), and the insets of (a). Bottom row: The doublon number $N_D$ in the ground eigenstate as a function of $\beta$, for the values of $U$ corresponding to the upper panels.}
    \label{fig:exact-eigenspectrum}
\end{figure}

To understand how the ground state changes with $\beta$, we start with a very small system, where we can diagonalize the Hamiltonian exactly and inspect the full eigenspectrum directly. We consider $L=8$ sites and population $N_\uparrow+N_\downarrow = 2+2$. For such a small system, we can diagonalize the Hamiltonian matrix to obtain the complete eigenspectrum for any given $U$ and $\beta$.

Before we start, let us briefly review the known eigenstates of the $U < 0$ Hubbard Hamiltonian in the homogeneous limit ($\beta = 0$), focusing on the balanced-population case (${N_\uparrow = N_\downarrow}$), for later comparison with $\beta > 0$. The Schr\"{o}dinger equation for this Hamiltonian can be solved via the Bethe ansatz approach and, for periodic boundary conditions, the solution shows that each eigenstate can be approximately interpreted as being made up of bound pairs alongside unpaired fermions~\cite{essler2005}. For large $|U|$, each bound pair is well approximated by a doublon (two opposite-spin fermions sitting on the same site), whereas for smaller $|U|$ the pairs become spatially extended. For any $U < 0$, the lowest-energy state is one in which all fermions are organized into bound pairs~\cite{essler2005}. In the following, we use these standard results to interpret the exact-diagonalization spectra of our small open-boundary chains.

In Fig.~\ref{fig:exact-eigenspectrum} (upper row), we show the eigenenergies obtained via exact diagonalization as functions of $\beta$ for different values of $U$. The energies are color-coded according to the number of doubly-occupied sites (the doublon number $N_D$) in each eigenstate, defined as
\begin{equation}
    N_D = \sum_{j=1}^L \langle \hat{n}_{j,D} \rangle,
\end{equation}
where the doublon density on site $j$ is given by
\begin{equation}
     \hat{n}_{j,D} = \hat{n}_{j,\uparrow} \hat{n}_{j,\downarrow}.
\end{equation}
The doublon number ranges from 0 to ${ N_D^\mathrm{max} = \min\left(N_\uparrow,N_\downarrow\right) }$. In the bottom row of Fig.~\ref{fig:exact-eigenspectrum}, we show $N_D$ in the ground eigenstate as a function of $\beta$, for the same values of $U$. Note that the doublon number is related to an eigenstate's total energy $E$ (by the Hellmann--Feynman theorem) as $N_D = \partial E/\partial U$.

\subsection{Strong-interaction case}

We first examine the limit of strong interactions, ${|U| \gg 1}$, shown in Fig.~\ref{fig:exact-eigenspectrum}(a) for $U = -15$. In this limit the hopping processes which change the doublon number are suppressed, and the Hamiltonian approximately conserves the total doublon number $N_D$. As a result, the eigenspectrum is organized into manifolds distinguished by approximately integer values of ${N_D \approx 0, 1, 2}$. These manifolds are separated by energies ${\sim |U|}$, while the spread (bandwidth) of levels within each manifold is set by the residual kinetic energy of doublons and unpaired fermions, and is at most of the order of the single-particle bandwidth (\emph{i.e.}, a few $t$). In this sense, the ``strong-coupling'' regime refers to $|U|$ being large compared to the kinetic-energy scale set by the single-particle bandwidth. Note that the precise threshold for strong coupling varies for lattices with different fillings and sizes.

At ${\beta = 0}$, each bound pair approximately constitutes a single doublon, so the manifolds can be interpreted as corresponding to different numbers of bound pairs vs. unpaired fermions. In principle, coincidental presence of two unpaired fermions on the same site also contributes to $N_D$, but in the strong-coupling limit such contributions are generally negligible compared to those from genuine bound pairs. At ${\beta=0}$, the ground state belongs to the ${N_D \approx 2}$ manifold.

Now let us consider what happens to these eigenstates as $\beta$ increases. The nearly doublonic pairs feel only a negligible net external potential, since the potentials felt by the two fermions on the same site cancel. It is therefore the unpaired fermions that dominate the $\beta$-dependence of the many-body energies. This $\beta$ dependence is visible in Fig.~\ref{fig:exact-eigenspectrum}(a) as a ``fanning out'' of levels within each manifold (except for the purely-doublon manifold). The spread of this ``fan'' grows with $\beta$, at a rate proportional to the number of unpaired fermions in the manifold. Within a given manifold, each eigenstate consists of $N_D$ bound pairs and ${N - 2N_D}$ unpaired fermions, which can occupy different combinations of single-particle orbitals. Since some orbitals increase in energy with $\beta$ while others decrease, different eigenstates within the same manifold show either an overall energy increase or decrease with $\beta$.

As $\beta$ increases, the ground state energy is repeatedly crossed by energies from lower-$N_D$ manifolds. Each such crossing decreases the ground-state $N_D$. In the ${U \to -\infty}$ limit, these are exact crossings, and the ground state switches discontinuously from one integer $N_D$ value to the next. For large but finite $|U|$, the hopping term weakly mixes the crossing eigenstates, turning each crossing into a narrow anticrossing. Consequently, with increasing $\beta$, the ground-state $N_D$ is seen to switch suddenly but continuously between near-integer plateaus. The two insets in Fig.~\ref{fig:exact-eigenspectrum}(a) (top) show magnified views of these anticrossings. In Fig.~\ref{fig:exact-eigenspectrum}(a) (bottom), the stepwise behavior of $N_D$ in the ground state is shown directly. It can be seen that $N_D$ switches between approximately integer values, as ${2 \to 1 \to 0}$. 

This behavior can be interpreted as the successive breaking of bound pairs, with each pair being replaced by two unpaired fermions. Physically, as $\beta$ grows, the external potential increasingly favors the spatial separation of spin components by lowering the energy of unpaired single-particle orbitals. When a particular threshold value of $\beta$ is reached, the potential overcomes the attractive binding and breaks another pair.

Overall, for strong $|U|$, these results paint a clear picture: as $\beta$ is increased, the external potential breaks the pairs \emph{one-by-one}. Therefore it is possible to precisely tune the number of bound pairs versus unpaired fermions.

\subsection{Moderate interaction case}

The picture becomes more complicated for weaker interactions, and in Fig.~\ref{fig:exact-eigenspectrum}(b), we show the situation for ${U = -2.5}$. At this moderate interaction strength, the hopping is strong enough to mix configurations with different $N_D$, and the manifolds with different doublon numbers melt into each other. At the same time, bound pairs are no longer well approximated by on-site doublons: a pair's two constituent fermions can be found with significant probability at different sites. Therefore each pair contributes less than one doublon on average. For example, at $\beta = 0$, the ground-state $N_D$ obtained from exact diagonalization is only $\approx 1.2$, even though the Bethe-ansatz picture for the homogeneous model predicts two bound pairs in the ground state.

At this value of $U$, increasing $\beta$ still causes the ground-state $N_D$ to decrease in a stepwise manner, indicating that bound pairs still exist as distinct correlation patterns. However, the avoided crossings are now broadened and overlapping, and the ground-state $N_D$ evolves smoothly rather than in sharp steps. Physically, one can say that, because the pairs are loosely bound, even small changes in $\beta$ can now modify a pair's internal structure. Therefore, instead of breaking abruptly at specific $\beta$ values, each pair now loosens gradually with increasing $\beta$.

Finally, to illustrate how the moderate-$U$ behavior connects to the non-interacting limit, we show the ${U = 0}$ case in Fig.~\ref{fig:exact-eigenspectrum}(c). In this limit the distinction between bound pairs and free fermions vanishes entirely. The nonzero $N_D$ in each eigenstate arises solely from coincidental finding of two unbound opposite-spin fermions on the same site, and the local double-occupancy is simply a product of single-particle densities: ${ \langle \hat{n}_{j,\uparrow} \hat{n}_{j,\downarrow} \rangle } = { \langle \hat{n}_{j,\uparrow} \rangle \langle \hat{n}_{j,\downarrow} \rangle }$. There is still a smooth $\beta$-dependence of $N_D$, which can be attributed entirely to the continuous deformation of single-particle orbitals. Specifically, the ground state at any $\beta$ is constructed by filling the lowest single-particle orbitals of both spin components. As $\beta$ grows, these lowest-energy orbitals are increasingly localized near opposite ends of the lattice, reducing the overlap between opposite spins.

\section{Numerical results for larger systems}
\label{sec:results-from-dmrg}

The previous section has given an intuitive picture of the ground-state properties as a function of $\beta$. In particular, we have shown that varying $\beta$ changes the ground-state doublon number $N_D$, which can be treated as a proxy for the number of bound pairs. To assess how well these observations scale to larger particle numbers, we next show the results of the DMRG analysis of larger systems. The lattice size is set to ${L = 120}$ sites, and throughout this section we focus on two balanced fillings: ${N = 10+10}$ (\emph{i.e.}, ${n = 1/6}$) and ${N = 45+45}$ (\emph{i.e.}, ${n = 3/4}$). Unless stated otherwise, henceforth $N_D$ will always be understood to refer to the ground state.

\subsection{DMRG numerical settings}
\label{sec:dmrg-settings}

We use DMRG to numerically find the ground state at varying $\beta$. The DMRG calculations require a careful choice of numerical parameters, such as the maximum bond dimension $\chi_\mathrm{max}$ and the total number of sweeps~\cite{10.21468/SciPostPhysLectNotes.5}. For our system, we find the following settings appropriate: maximum bond dimension ${\chi_\mathrm{max} = 400}$, with up to 140 sweeps allowed (in practice DMRG converges in far fewer sweeps). We found that, for certain ranges of $\beta$, DMRG can converge to different eigenstates, depending on the initial state used to seed the variational optimization. This sensitivity appears at strong $|U|$, where the number of bound pairs is approximately conserved, but not at weak $|U|$, where the Hamiltonian strongly couples states with different $N_D$. We therefore attribute it to the DMRG optimization becoming trapped in a sector with a fixed number of bound pairs. To mitigate this effect, for each value of $\beta$ we have run DMRG ${1+N/2}$ times, starting from ${1+N/2}$ different initial seeds (configurations with different doublon numbers, so as to overlap with eigenstates from different ${ N_D=0,1,\ldots,N/2 }$ manifolds), and we selected the final state with the lowest converged energy.

We have checked that the above approach yields converged results for all the parameter regimes included in our figures (\emph{i.e.}, the observables plotted in the figures do not change when $\chi_\mathrm{max}$ is increased). The discarded weight is $< 10^{-7}$ in the final sweep. We have also checked that the energy variance ${\langle \hat{H}^2 \rangle - \langle \hat{H} \rangle^2}$ is negligible across all tested parameter ranges, indicating that in each case DMRG converges correctly to an eigenstate.

\subsection{Number of doublons depending on $\beta$}
\label{sec:results-doublon-number}

\begin{figure}[t]
    \centering
    \includegraphics[width=0.95\linewidth]{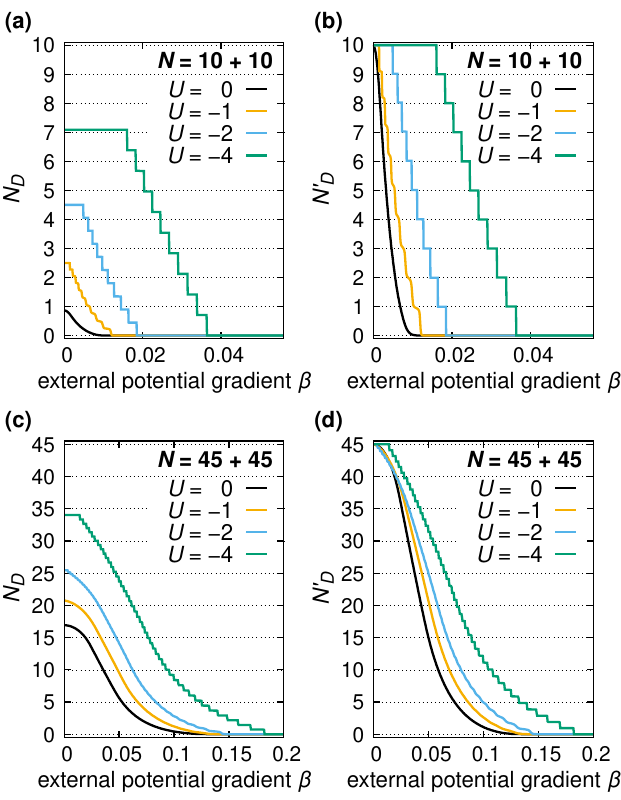}
    \caption{(a) The ground-state doublon number $N_D$, as a function of $U$ and $\beta$, for $L=120$ and $N_\uparrow+N_\downarrow=10+10$. (b) The corresponding rescaled doublon number $N'_D$ [Eq.~\eqref{eq:rescaled_doublon_number}], which approximates the total number of bound pairs. (c,d) $N_D$ and $N'_D$ for $L=120$ and $N=45+45$.}
    \label{fig:doublon-numbers-larger-systems}
\end{figure}

In Fig.~\ref{fig:doublon-numbers-larger-systems}, we show the ground-state $N_D$ as a function of $\beta$, for several values of $U$ and two populations: ${N=10+10}$ [Fig.~\ref{fig:doublon-numbers-larger-systems}(a)] and ${N=45+45}$ [Fig.~\ref{fig:doublon-numbers-larger-systems}(c)]. We also show the corresponding \emph{rescaled} doublon count $N'_D$ in Fig.~\ref{fig:doublon-numbers-larger-systems}(b,d). This quantity is defined as
\begin{equation}
    \label{eq:rescaled_doublon_number}
    N'_D = \frac{N}{2} \frac{N_D}{N_D(\beta=0)}
\end{equation}
and serves as a proxy estimate for the number of bound pairs in the system. This estimate assumes that the number of bound pairs at ${\beta=0}$ is $N/2$ (as in the usual ${\beta=0}$ ground state), and that the number of doublons per bound pair remains constant as $\beta$ increases. The rescaled $N'_D$ allows us to directly compare results across different values of $U$ and $N$. For moderate and strong attraction it provides a reasonable estimate of the effective number of bound pairs, whereas in the weak-coupling regime it should be interpreted more cautiously. Later, in Sec.~\ref{sec:pair-counting-from-excess-density}, we demonstrate an alternative bound pair counting quantity, which is based solely on one-body densities and makes no reference to the number of doublons per pair.

First let us look at the ${N=10+10}$ case [Fig.~\ref{fig:doublon-numbers-larger-systems}(a,b)]. The behavior closely mirrors that of the ${2+2}$ system: as $\beta$ increases, $N_D$ decreases through a series of plateaus in a stepwise fashion. Each plateau corresponds to an approximately integer value of $N'_D$, which suggests that the change of ground state under increasing $\beta$ can still be described as one-by-one breaking of bound pairs. As $|U|$ is decreased, the plateaus gradually soften, analogously to what we observed in the ${N=2+2}$ spectrum.

For the higher particle number ${N=45+45}$ [Fig.~\ref{fig:doublon-numbers-larger-systems}(c,d)], $N_D$ behaves similarly. In particular, at $U=-4$, we resolve a sequence of plateaus in $N'_D$ with values close to successive integers ${45, 44, \ldots, 0}$, indicating that the number of bound pairs can be controlled with high relative precision.

The assumption that the number of doublons per pair is independent of $\beta$ is only approximate, since the increasing external potential gradient is expected to loosen the pairs somewhat. This loosening shows up as a smoothing of the plateaus, but it is worth checking that the discrete drops in $N_D$ are not themselves caused by an abrupt loosening of the pairs, rather than by pair breaking. In Sec.~\ref{sec:pair-loosening}, we explore pair loosening quantitatively for $|U|$ at which $N'_D$ shows clear plateaus. We show that the drops cannot be ascribed to pair loosening, confirming that genuine pair breaking is the mechanism behind them.

\subsection{Effect of gradient $\beta$ on the density profiles}
\label{sec:results-density-profiles}

\begin{figure}[t]
    \centering
    \includegraphics[width=0.95\linewidth]{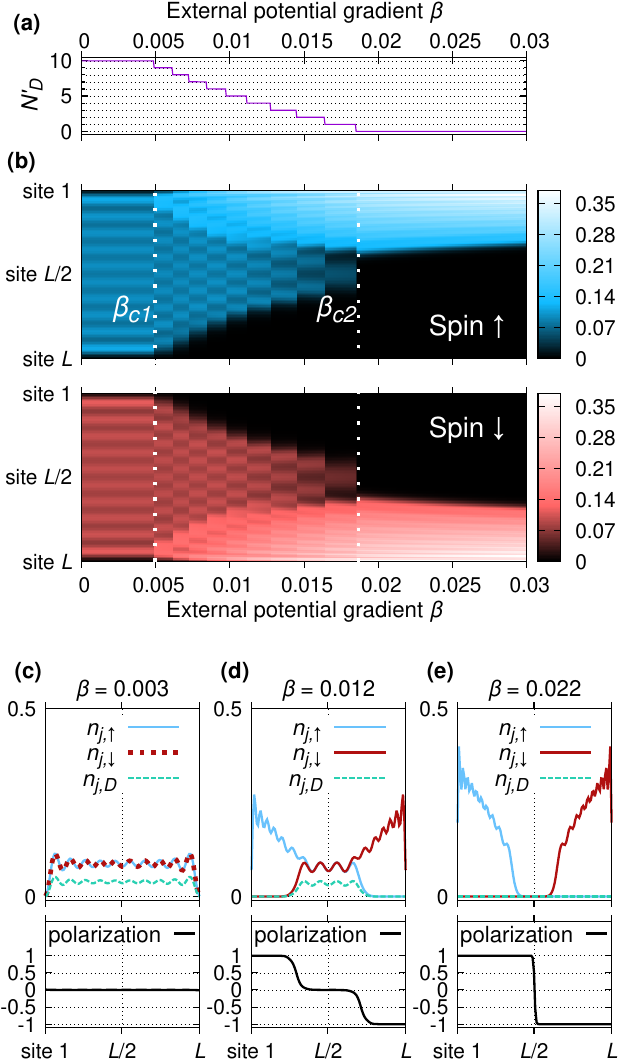}
    \caption{(a) Ground-state rescaled doublon number $N'_D$ as function of $\beta$, for an example system with $L=120$, $U=-2$, $N=10+10$. (b) Ground-state density profiles $n_{j,\uparrow} = {\langle \hat{n}_{j,\uparrow} \rangle}$ (upper) and $n_{j,\downarrow} = {\langle \hat{n}_{j,\downarrow} \rangle}$ (lower) at sites $j = 1, \ldots, L$ as a function of $\beta$, for the same system. The local density on each site is indicated by a color scale. The two dotted lines mark the critical gradient values, $\beta_{c1}$ and $\beta_{c2}$, at which the ground-state density undergoes a sudden change in behavior. (c, d, e) Example density profiles for three different values of $\beta$, representing: (c) the small-gradient regime $\beta < \beta_{c1}$, (d) the intermediate-gradient regime $\beta_{c1} \le \beta < \beta_{c2}$, and (e) the large-gradient regime $\beta_{c2} \le \beta$. The value of $\beta$ is indicated above each plot. The bottom plots show the local relative spin polarization $p_j = { ( n_{j,\uparrow} - n_{j,\downarrow} ) / ( n_{j,\uparrow} + n_{j,\downarrow}) }$. In plot (c), $n_{j,\downarrow}$ is plotted as a dotted line in order to more clearly show that $n_{j,\downarrow} \approx  n_{j,\uparrow}$ everywhere.}
    \label{fig:ground-state-density-change-dmrg}
\end{figure}

\begin{figure}[t]
    \centering
    \includegraphics[width=0.95\linewidth]{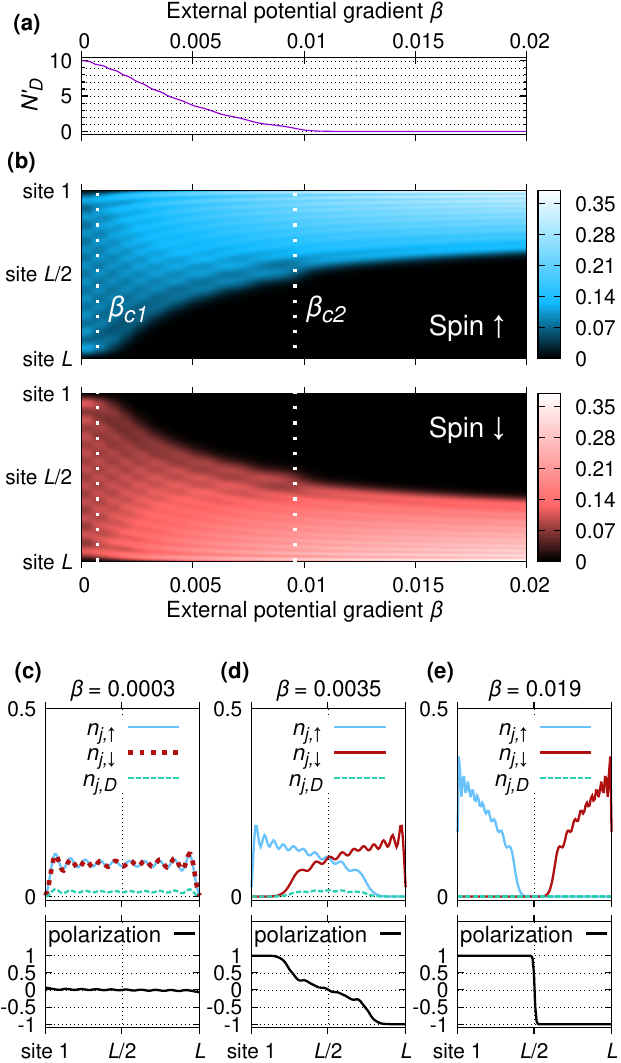}
    \caption{Same as Fig.~\ref{fig:ground-state-density-change-dmrg}, but for $U = -0.5$.}
    \label{fig:ground-state-density-change-dmrg-2}
\end{figure}

We now examine the ground states more closely by analyzing the density distribution of the spin components across the lattice. Fig.~\ref{fig:ground-state-density-change-dmrg} provides an overview of how the ground state depends on $\beta$, with a focus on the one-body density profiles $n_{j,\sigma} = {\langle \hat{n}_{j,\sigma} \rangle}$ and the doublon density profile $n_{j,D} = \langle \hat{n}_{j,D} \rangle$. The system parameters are $U = -2$ and $N=10+10$ ($n=1/6$). Fig.~\ref{fig:ground-state-density-change-dmrg}(a) shows how $N'_D$ in the ground state changes as $\beta$ is varied, while the density plots in Fig.~\ref{fig:ground-state-density-change-dmrg}(b) depict the accompanying changes in density profiles of both spin components. For three representative values of $\beta$, the density profiles are shown in detail in Fig.~\ref{fig:ground-state-density-change-dmrg}(c-e), together with the local relative spin polarization $p_j = { ( n_{j,\uparrow} - n_{j,\downarrow} ) / ( n_{j,\uparrow} + n_{j,\downarrow} ) }$. Note that, due to the symmetries of the Hamiltonian, the density profile is always symmetric under simultaneous spatial reflection and spin exchange: $ n_{j,\uparrow} = n_{L-j+1,\downarrow}$.

One can clearly see that each sharp change in the $N'_D$ value is accompanied by a sharp change in the density distributions. To characterize this behavior, we define two threshold gradients: $\beta_{c1}$, at which the first pair breaks, and $\beta_{c2}$, at which the final pair breaks. For the $U=-2$ example shown here, these thresholds can be visually estimated from Fig.~\ref{fig:ground-state-density-change-dmrg}(a) as $\beta_{c1} \approx 0.005$ and $\beta_{c2} \approx 0.0185$, and are indicated by dotted lines in Fig.~\ref{fig:ground-state-density-change-dmrg}(b). These thresholds delineate three qualitatively distinct regimes:

\begin{itemize}

\item $\beta < \beta_{c1}$ (no pairs broken, $N'_D = N/2$): All sites are unpolarized ($p_j \approx 0$). The density is uniform across the bulk of the lattice, up to Friedel-like standing-wave oscillations and the vanishing of the density at the hard edges, both induced by the open boundaries. An exemplary density profile from this regime is shown in Fig.~\ref{fig:ground-state-density-change-dmrg}(c) for $\beta = 0.003$. In this regime, the density profile is nearly insensitive to $\beta$, indicating that the external potential is still too weak to overcome the attraction.

\item $\beta_{c1} \le \beta < \beta_{c2}$ (successive pair-breaking, ${0 < N'_D < N/2}$): This is the regime where the external potential is strong enough to overcome the attraction. As pair-breaking starts, phase separation emerges. One can discern a paired, unpolarized core ($p_j \approx 0$, $n_{j,D} > 0$), and surrounding fully polarized, unpaired wings of opposite spins on opposite lattice edges ($p_j = \pm 1$, $n_{j,D} = 0$). The core and the wings are separated by partially polarized transitional regions ($0 < |p_j| < 1$), whose width depends on the parameters. With each drop in $N'_D$, the paired core shrinks. This reflects a pair disappearing and being replaced by two unpaired fermions sitting in the opposite wings. The density thus directly reflects how the external potential draws the two spin components to opposite ends. An exemplary density profile from this regime of $\beta$ is shown in Fig.~\ref{fig:ground-state-density-change-dmrg}(d) for $\beta = 0.012$. 

\item $\beta \ge \beta_{c2}$ (all pairs broken, $N'_D = 0$): The density is now fully spin-separated by the external potential, with the two spin components localized at opposite ends of the lattice. An exemplary density profile from this regime of $\beta$ is shown in Fig.~\ref{fig:ground-state-density-change-dmrg}(e) for ${\beta = 0.022}$. Further increases in $\beta$ only reduce the width of the wings, compressing the density profile closer to the lattice edges.

\end{itemize}

Fig.~\ref{fig:ground-state-density-change-dmrg-2} shows an example for a weaker interaction, $U = -0.5$, for which pair breaking occurs smoothly, over a broad range of $\beta$. As a result, the $\beta$-dependency of the density distribution is smoother, and the critical thresholds are less sharply defined (the dotted lines show rough visual estimates: $\beta_{c1} \approx 0.0005$, $\beta_{c2} \approx 0.0098$). An additional effect is that the central core region displays partial polarization throughout, rather than being uniformly unpolarized. Nevertheless, we can still approximately distinguish three different ranges of $\beta$ with qualitatively different ground-state density distributions.

\subsection{Pair counting via one-body densities}
\label{sec:pair-counting-from-excess-density}

As we have shown in Sec.~\ref{sec:results-density-profiles}, pair breaking is already visible at the level of the one-body densities, as a dipole-like spin polarization across the lattice. This suggests an alternative way of counting pairs via one-body observables only.

We define the excess population $N^{\mathrm{ex}}_\sigma$ as the total number of spin-$\sigma$ fermions that have left the unpolarized core and generated a local spin imbalance elsewhere:
\begin{equation}
    \label{eq:excess}
N^{\mathrm{ex}}_\sigma = \sum_i \max\left(\langle \hat{n}_{i,\sigma} - \hat{n}_{i,-\sigma} \rangle,0\right).
\end{equation}
The number of bound pairs can then be estimated as
\begin{equation}
\label{eq:pair_counter_from_excess}
    \min\left( N_\uparrow - N_\uparrow^{\mathrm{ex}},\; N_\downarrow - N_\downarrow^{\mathrm{ex}} \right).
\end{equation}
Equivalently, $N^{\mathrm{ex}}_\sigma$ can be described as the population of a polarized ``wing'' region near the lattice edge, properly including the partially-polarized transition region between the wing and the core.

The physical motivation is the following. In a uniform system with a global spin imbalance $N_\uparrow \ne N_\downarrow$, at least $\left|N_\uparrow-N_\downarrow\right|$ majority fermions must remain unpaired. The definition of $N^{\mathrm{ex}}_\sigma$ is a local-density-approximation--based extension of this counting argument to nonuniform systems, which allows one to estimate the total number of unpaired fermions. Note that ${N^{\mathrm{ex}}_\uparrow - N^{\mathrm{ex}}_\downarrow} = {\sum_i \langle \hat{n}_{i,\uparrow} - \hat{n}_{i,\downarrow} \rangle} = {N_\uparrow-N_\downarrow}$, so in a balanced system the two excesses coincide: $N^{\mathrm{ex}}_\uparrow = N^{\mathrm{ex}}_\downarrow$.

\begin{figure}
    \centering
    \includegraphics[width=0.96\linewidth]{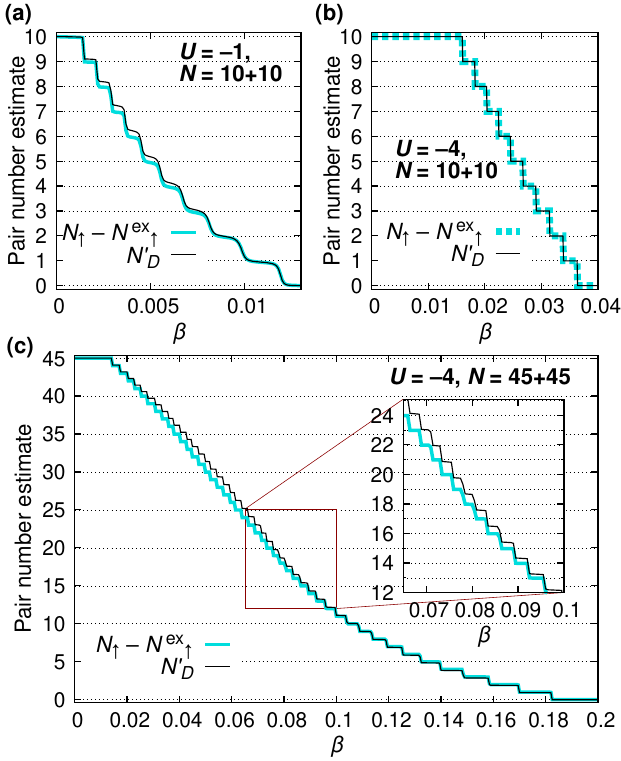}
    \caption{The $N_\uparrow - N^{\mathrm{ex}}_\uparrow$ estimate for the number of bound pairs (thick line) and the pair-counting quantity $N'_D$ (thin line), for ${L=120}$ systems with (a) $U=-1$, $N=10+10$, (b) $U=-4$, $N=10+10$, and (c) $U=-4$, $N=45+45$. The inset in (c) zooms in on a range of $\beta$ to show discrepancies between both estimators.}
    \label{fig:wing_population}
\end{figure}

In Fig.~\ref{fig:wing_population} we plot ${N_\uparrow - N^{\mathrm{ex}}_\uparrow}$ as a function of $\beta$ and compare it with $N'_D$. The three parameter sets represent three distinct cases: dilute systems at moderate and strong interactions [Fig.~\ref{fig:wing_population}(a): $N=10+10$, $U = -1$; Fig.~\ref{fig:wing_population}(b): $N=10+10$, $U = -4$], and a dense system [Fig.~\ref{fig:wing_population}(c): $N=45+45$, $U=-4$]. In all three cases, ${N_\uparrow - N^{\mathrm{ex}}_\uparrow}$ forms clear integer plateaus, and the pair number it predicts is consistent with $N'_D$. The only significant discrepancy occurs for the large population [Fig.~\ref{fig:wing_population}(c)], where the two estimators can differ by up to one pair (see the inset). For a system of $45+45$ fermions, however, this discrepancy in pair count amounts to a negligible relative error of $\sim 2\%$.

For completeness, we note that bound pairs can in principle contribute to $N^{\mathrm{ex}}_\sigma$ as well. If a pair is not strictly doublonic but extended over several sites, the spin-dependent potential can give it a nonzero dipole moment, displacing the two opposite-spin fermions to either side of the pair center of mass. Such a stretched pair would then shift the local spin imbalance slightly and add a fractional contribution to both $N^{\mathrm{ex}}_\uparrow$ and $N^{\mathrm{ex}}_\downarrow$. However, such fractional contributions are likely minimal, since Fig.~\ref{fig:wing_population} shows that the $N_\uparrow - N^{\mathrm{ex}}_\uparrow$ plateaus are always at approximately integer values.

We conclude that ${N_\uparrow - N^{\mathrm{ex}}_\uparrow}$ can be considered a good estimator of the pair number. Notably, unlike for $N'_D$, where the integer plateaus appear once the $N_D(\beta=0)$ value is explicitly rescaled to equal $N/2$, for the ${N_\uparrow - N^{\mathrm{ex}}_\uparrow}$ estimator we obtain integer plateaus without any rescaling, and without any assumption as to the number of doublons per pair. Compared to $N'_D$, this estimator has the advantage of being accessible from one-body measurements alone, without reference to two-body correlations, which is an important consideration for experimental implementations.

\subsection{Dependency of $\beta_{c1},\beta_{c2}$ on system parameters}
\label{sec:dependency-of-critical-beta-on-parameters}

It is useful to establish how $\beta_{c1}$ and $\beta_{c2}$ depend on the parameters $U$, $n$, and $L$. We will first examine this dependence qualitatively, using DMRG numerical results. Later, in Sec.~\ref{sec:phenomenological-model} and Sec.~\ref{sec:lda}, we will find approximate analytical expressions.

\begin{figure}[t]
    \centering
    \includegraphics[width=0.95\linewidth]{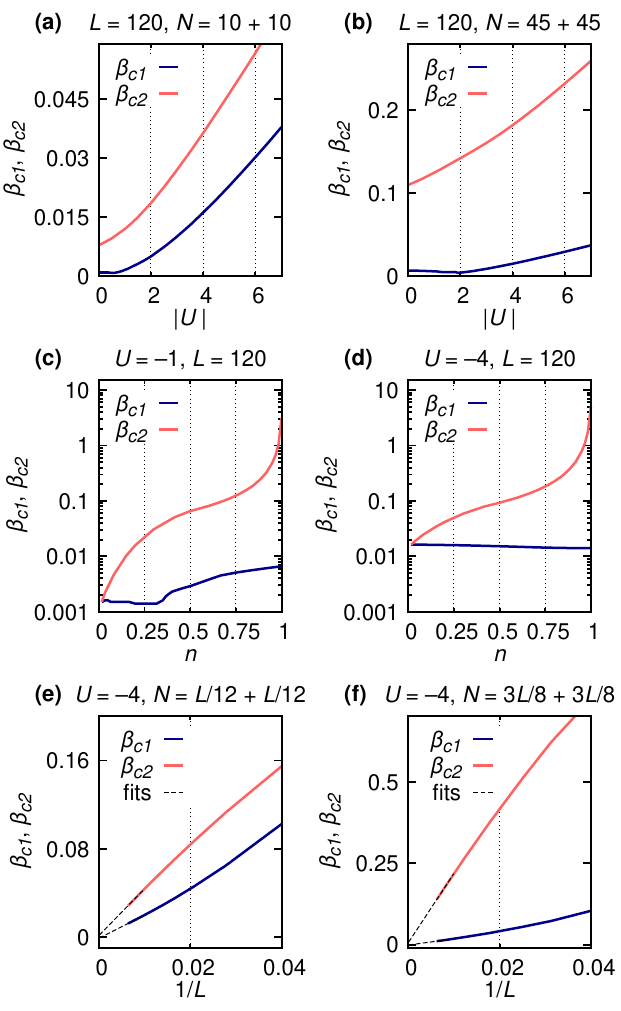} 
    \caption{(a,b) Threshold values $\beta_{c1}$ and $\beta_{c2}$ as a function of $U$, for $L=120$ and (a) $N=10+10$ or (b) $N=45+45$. (c,d) Same, as a function of $n$, for a system with $L=120$ and (c) $U=-1$ or (d) $U=-4$; note the logarithmic scale. (e,f) Same, as a function of $1/L$, for a system with $U = -4$ and (e) $N=L/12+L/12$ or (f) $N=3L/8+3L/8$. The dashed lines indicate fits to $\mathrm{const}_1 L^{-1} + \mathrm{const}_2$, fitted in the range $1/L \le 0.01$; these fits show that $\beta_{c1}$, $\beta_{c2}$ reach 0 in the $1/L \to 0$ limit.}
    \label{fig:critical-beta-values}
\end{figure}

In Fig.~\ref{fig:critical-beta-values}(a,b), we show $\beta_{c1}$ and $\beta_{c2}$ as functions of $U$ for $L=120$ and two constant fillings: (a) $N=10+10$, (b) $N=45+45$. The threshold values are determined as follows: $\beta_{c1}$ is the smallest value of $\beta$ for which $N'_D(\beta) \le N/2 - 1/2$ (\emph{i.e.}, the midpoint between zero and one broken pairs), and $\beta_{c2}$ is the smallest $\beta$ for which $N'_D(\beta) \le 1/2$. These definitions provide an operational way to extract $\beta_{c1}$ and $\beta_{c2}$ from numerical data even when the individual plateaus in $N'_D$ are not perfectly sharp. Of course, for $U=0$ the threshold values become meaningless physically as there are no pairs to break, but $\beta_{c1}$ and $\beta_{c2}$ vary smoothly all the way to that limit.

One can see that $\beta_{c2}$ is a monotonically increasing function of $|U|$. The behavior of $\beta_{c1}$ is more nuanced: for weaker interactions it depends only weakly on $|U|$, as is particularly visible for the larger population in Fig.~\ref{fig:critical-beta-values}(b), but at sufficiently large $|U|$ it too increases monotonically. This is intuitive, as stronger attraction means that breaking each pair---whether the first or the last---requires a larger potential gradient, \emph{i.e.}, a larger potential difference between the center of the lattice and the potential minimum.

Next, we consider the dependence of $\beta_{c1}$ and $\beta_{c2}$ on $n$. In Fig.~\ref{fig:critical-beta-values}(c,d), we show these values for varying $n$ at fixed $L=120$ and (c) $U = -1$, (d) $U = -4$. One can see that the threshold $\beta_{c2}$ is highly sensitive to the population; it increases monotonically with $n$ and varies by several orders of magnitude over the range $n = 0 \to 1$. This monotonic increase can be intuitively understood: larger populations contain more pairs to break, so a larger $\beta$ is generally needed to break the last one. By contrast, $\beta_{c1}$ depends much more weakly on $n$. For $U = -1$, it changes by less than an order of magnitude across the same range, while for $U = -4$ it is nearly constant. This suggests that the first pair-breaking event is largely insensitive to the presence of other pairs, especially at stronger interactions where pairs behave as localized doublons with very little mutual interaction.

In the $n \to 0$ limit the values $\beta_{c1}$ and $\beta_{c2}$ become equal, since in the extreme $N=1+1$ case there is only one pair to break. At the opposite end, for $n > 1$ (over half filling), simple particle-number counting shows that at least one site must be doubly occupied in any lattice configuration. Consequently, $N_D$ cannot be reduced to zero and our operational definition of $\beta_{c2}$ ceases to be meaningful; this is reflected in $\beta_{c2}$ tending to very large values as $n \to 1$.

Finally, we consider the dependence of $\beta_{c1}$ and $\beta_{c2}$ on $L$. In Fig.~\ref{fig:critical-beta-values}(e,f), we show these values as functions of $1/L$ for $U = -4$ at constant filling: (e) $N=L/12+L/12$ and (f) $N=3L/8+3L/8$. Both threshold values increase monotonically with $1/L$. For large $L$, they become linear in $1/L$ and approach zero as $1/L \to 0$. This is demonstrated by the dashed lines in Fig.~\ref{fig:critical-beta-values}(e,f), which show the linear fits of $\beta_{c1}$ and $\beta_{c2}$ in the range $1/L \le 0.01$. This behavior has a simple physical explanation: for larger $L$, a given gradient $\beta$ produces a larger potential difference between the ends of the lattice, and the energy gain from separating the two spin components grows as $\sim \beta L$. In the infinite-system limit ($L\to\infty$ at fixed $n, U, \beta$), any nonzero gradient favors complete spin separation, so $\beta_{c1}, \beta_{c2} \to 0$. Consequently, for larger $L$, finer control of $\beta$ is required to precisely tune the number of bound pairs.

\section{Pair loosening under increasing $\beta$}
\label{sec:pair-loosening}

So far, we have assumed that for moderate $|U|$---where $N_D$ decreases in sharp discrete steps---the decrease in the doublon number $N_D$ is due solely to the breaking of bound pairs. This assumption, however, deserves closer scrutiny. In principle, the number of pairs in the range $\beta_{c1} < \beta < \beta_{c2}$ might not decrease at all; instead, each drop in $N_D$ could correspond to an abrupt \emph{loosening} of the existing pairs, so that the number of doublons per pair is reduced. Such behavior is plausible on intuitive grounds, since increasing $\beta$ pulls the opposite-spin fermions of a pair in opposite directions and we would therefore expect it to stretch the pair out. The fact that the rescaled $N'_D$ steps fall almost exactly at integer values argues against this scenario, but does not exclude it.

Assessing this possibility is not simple, because the number of doublons per bound pair is not straightforward to determine. In this section we therefore address the objection from a different angle, by analyzing the two-body spatial correlations between opposite spins. We establish how strongly the pair extension changes as $\beta$ increases, and whether it is plausible that the $N_D$ drops are generated primarily by pair loosening.

We characterize the pairing correlations through the connected density-density correlator
\begin{equation}
    \label{eq:connected-density-density-correlator}
    G_{ij} = \langle \hat{n}_{i,\downarrow} \hat{n}_{j,\uparrow} \rangle - \langle \hat{n}_{i,\downarrow} \rangle \langle \hat{n}_{j,\uparrow} \rangle.
\end{equation}
This quantity captures two-body correlations between opposite spins while subtracting the background of coincidental correlations already contained in the one-body densities. In the non-interacting limit $U=0$, $G_{ij}$ vanishes identically for any $\beta$, and it acquires nonzero features only at finite interaction strength. It therefore isolates precisely those features that cannot be explained at the one-body level and that must originate from the interaction term of the Hamiltonian. Note also that, because $N_\uparrow$ and $N_\downarrow$ are fixed, each row of $G_{ij}$ obeys the sum rule $\sum_j G_{ij} = 0$, so that every positive region of $G$ is compensated by a negative one.

\begin{figure}
    \centering
    \includegraphics[width=0.96\linewidth]{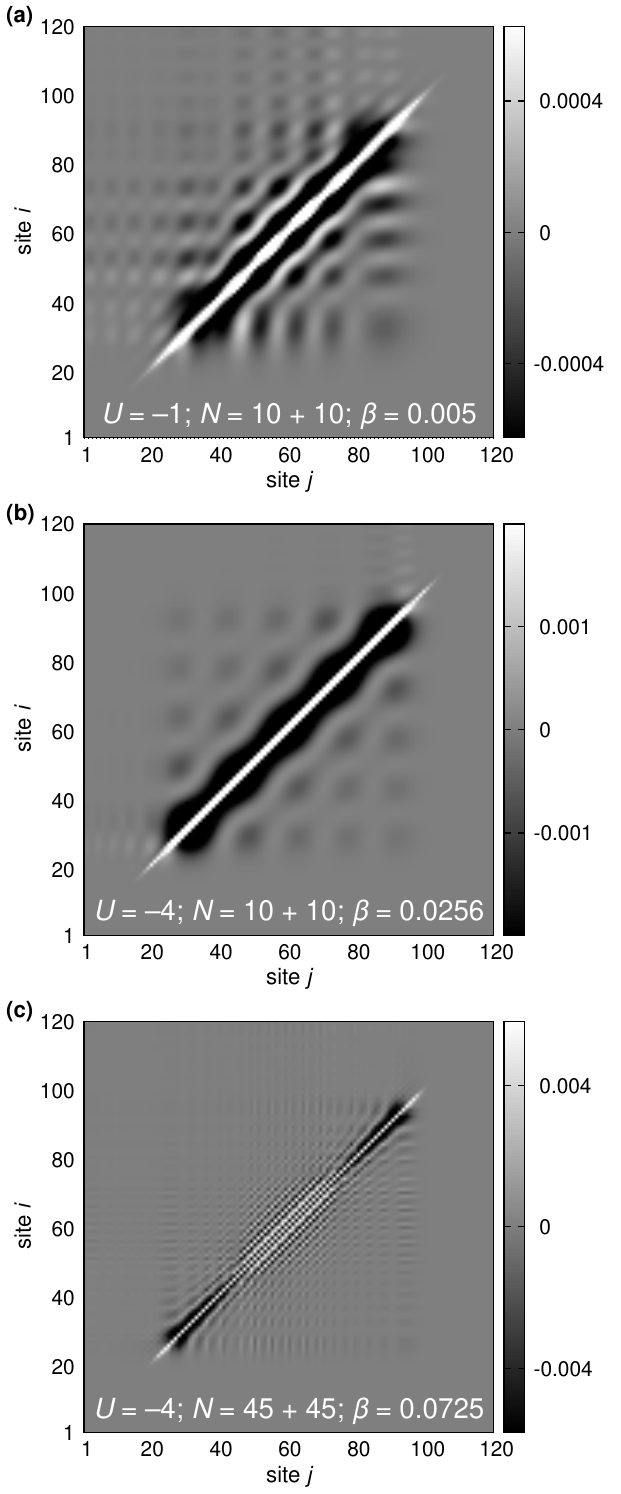}
    \caption{The connected correlator [Eq.~\eqref{eq:connected-density-density-correlator}] for $L=120$ systems with three example parameter sets: (a) $L=120, N=10+10, U = -1, \beta = 0.005$; (b) $L=120, N=10+10, U = -4, \beta = 0.0256$; (c) $L=120, N=45+45, U = -4, \beta = 0.0725$. In each case $\beta$ is chosen to fall in the middle of a plateau with $N'_D \approx 5$ ($N=10+10$) or $N'_D \approx 22$ ($N=45+45$). The color scale is adjusted in each panel so as to reveal the weak structure: the scale runs between the values ${\pm 0.04\,G_{L/2,L/2} }$, and the mid-gray shade always corresponds to zero.}
    \label{fig:two-body-up-down-spin-correlations}
\end{figure}

In Fig.~\ref{fig:two-body-up-down-spin-correlations} we plot $G_{ij}$ for ${L=120}$ and three parameter sets: (a) $U=-1$, $N=10+10$, (b) $U=-4$, $N=10+10$, and (c) $U=-4$, $N=45+45$. In each case we chose $\beta$ as the value corresponding to an $N'_D$ plateau with $N'_D \approx N/4$, so that the plots show a representative picture of the pairing within the $\beta_{c1} < \beta < \beta_{c2}$ regime.

The structure of $G_{ij}$ can be separated, roughly, into two main contributions: intra-pair correlations---between the opposite-spin fermions that form a single bound pair---and inter-pair correlations, between opposite-spin fermions belonging to distinct pairs. The intra-pair correlations appear as a region of positive $G_{ij}$ surrounding the diagonal $i=j$, and how far this region extends from the diagonal roughly reflects the extension of the bound pairs. The inter-pair correlations appear as regularly spaced oscillations between negative and positive $G_{ij}$, parallel to the diagonal. Physically, these regular oscillations reflect the fact that, in the strong-attraction limit, bound pairs behave as hardcore bosons with effective nearest-neighbor repulsion~\cite{RevModPhys.62.113} and therefore tend to maintain a preferred mutual separation. As a result, the probability of finding two opposite-spin fermions a given distance apart is enhanced at multiples of this separation and suppressed in between. The strongest connected correlations are present in the core region, away from the lattice edges. However, faint checkerboard-like oscillations can also be found in and near the wing regions (around $j \approx 1$ or $i \approx 120$); they are most discernible in Fig.~\ref{fig:two-body-up-down-spin-correlations}(a) ($U=-1$). These are residual correlations between the particle-number fluctuations of the standing-wave (Friedel-like) density patterns in the wings and in the core; they are two orders of magnitude weaker than the pairing correlations near the diagonal. These checkerboard features lack the structure that pairing would produce --- a positive ridge along ${j - i \approx \mathrm{const}}$ --- so they do not indicate pairing between core and wing fermions.

For a better overview of the correlations, we define the quantity $G(r)$, which sums the $\uparrow$--$\downarrow$ correlations at a fixed separation $j-i = r$ over all sites $i$:
\begin{equation}
    \label{eq:off-diagonal-correlator-trace}
G(r) = \sum_{i,j} G_{ij} \delta_{j,i+r}.
\end{equation}

\begin{figure}
    \centering
    \includegraphics[width=0.97\linewidth]{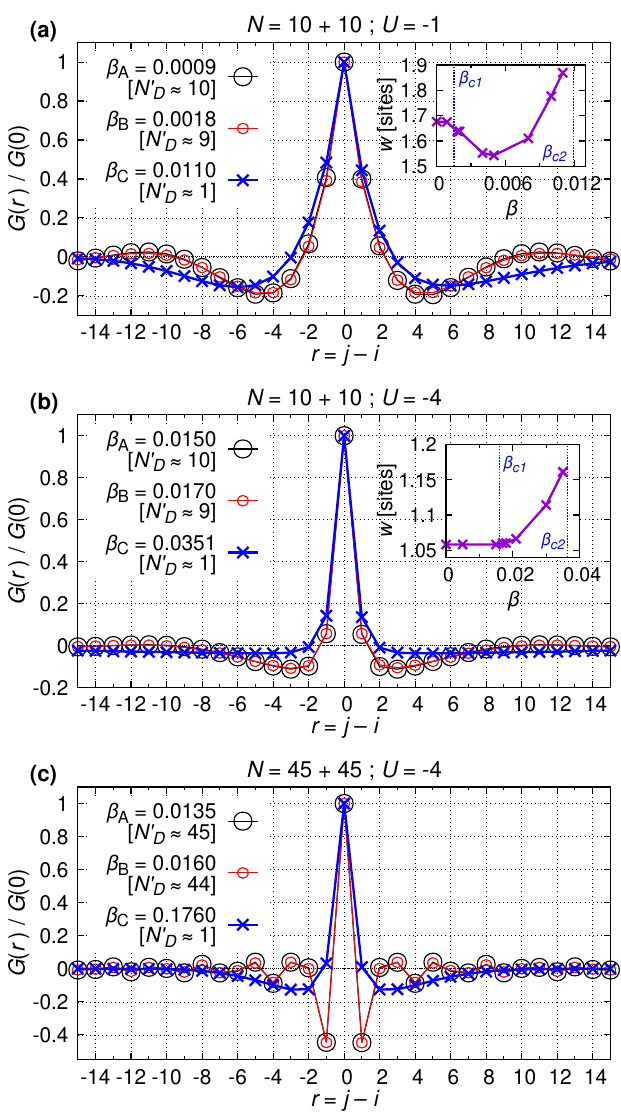}
    \caption{The integrated connected correlator $G(r)/G(0)$ [Eq.~\eqref{eq:off-diagonal-correlator-trace}] for several sets of system parameters and values of $\beta$. The peaks near $r=0$, with positive $G(r)$, correspond roughly to the profiles of the average bound pair. Insets in (a,b): the pair width $w$ [Eq.~\eqref{eq:full-width-at-half-maximum}], which quantifies the pair broadening, as a function of $\beta$; dashed lines indicate $\beta_{c1}$ and $\beta_{c2}$ for the given parameters. Crosses mark the calculated values of $w$, and the connecting lines are a guide to the eye.}
    \label{fig:quantity_Gr_averaging_over_correlator}
\end{figure}

In Fig.~\ref{fig:quantity_Gr_averaging_over_correlator} we show $G(r)/G(0)$ for the same $L$, $U$, and $N$ parameter sets as in Fig.~\ref{fig:two-body-up-down-spin-correlations}, in each case for three representative values of $\beta$: $\beta_A$, which lies just below $\beta_{c1}$; $\beta_B > \beta_{c1}$, taken in the middle of the $N'_D = N/2-1$ plateau, \emph{i.e.} just above $\beta_{c1}$; and $\beta_C < \beta_{c2}$, taken in the middle of the $N'_D = 1$ plateau. Normalizing by $G(0)$ removes any reduction in the overall correlation that stems purely from breaking the bound pairs, as opposed to pair loosening which would change (broaden) the profile of $G(r)$.

The averaged intra-pair correlations around the diagonal are reflected in $G(r)$ as a peak of positive $G(r)$, centered on $r=0$. Further away from $r=0$, $G(r)$ contains decaying oscillations which are the average over inter-pair correlations. Changes in the width of the peak at $r=0$ can be treated as an indicator of relative increase of pair extent. We stress, however, that $G(r)$ does not directly represent the actual two-body relative-coordinate wavefunction of a bound pair.

Comparing $U=-1$ [Fig.~\ref{fig:quantity_Gr_averaging_over_correlator}(a)] with $U=-4$ [Fig.~\ref{fig:quantity_Gr_averaging_over_correlator}(b,c)], the peak in $G(r)$ is narrower for the stronger attraction, as expected, since the pairs are more tightly bound at larger $|U|$. Comparing the curves for $\beta_A$ and $\beta_B$ (lines with large and small circles), we see that the intra-pair correlations are almost unchanged as $\beta_{c1}$ is crossed. At the larger value $\beta_C$, the intra-pair correlations do extend noticeably; even then, however, the peak of $G(r)$ spans only a few sites.

To quantify how much the pairs loosen within the range $\beta_{c1} < \beta < \beta_{c2}$, we characterize the intra-pair peak by its full width at half maximum $w$, defined as
\begin{align}
    w^+ = \min(r) &: \left(r>0, G(r)=\frac{1}{2}G(0)\right), \nonumber \\
    w^- = \min(|r|) &: \left(r<0, G(r)=\frac{1}{2}G(0)\right), \nonumber \\
    w = w^+ &+ w^-. \label{eq:full-width-at-half-maximum}
\end{align}
Here $w^+$ and $w^-$ are allowed to take non-integer values, with $G(r)$ at non-integer $r$ obtained by linear interpolation between adjacent $r$. Summing the two removes the slight asymmetry between $w^+$ and $w^-$ that is to be expected in a spin-dependent tilted potential. The change in the width $w$ quantifies how far the average pair is stretched as $\beta$ increases. Note, however, that it cannot be applied to the $N=45+45$ case [Fig.~\ref{fig:quantity_Gr_averaging_over_correlator}(c)]: in that case, at $\beta_A$ we find negative $G(r)$ at $r = \pm 1$ already, so that $w$ cannot meaningfully characterize the extension of a pair beyond the statement $w<1$.

The insets of Fig.~\ref{fig:quantity_Gr_averaging_over_correlator}(a) and (b) show $w$ as a function of $\beta$ for the corresponding parameter sets, with $\beta_{c1}$ and $\beta_{c2}$ marked by dashed lines. For $U=-1$ [Fig.~\ref{fig:quantity_Gr_averaging_over_correlator}(a)], crossing $\beta_{c1}$ \emph{reduces} $w$ slightly, from $\approx 1.68$ to $\approx 1.64$ sites ($2\%$ reduction). This is likely because the modifications to the inter-pair correlations slightly modify $G(r)$ near $r=0$. As we increase $\beta$ further across the full range $\beta_{c1} < \beta < \beta_{c2}$, $w$ grows from $\approx 1.64$ to $\approx 1.87$ sites, a broadening of $14\%$. A change this modest cannot account for the accompanying drop in the doublon number: the tenfold reduction from $N'_D=10$ to $N'_D=1$ would require each pair profile to broaden by a factor of $\sim 10$. The results for $U=-4$ [Fig.~\ref{fig:quantity_Gr_averaging_over_correlator}(b)] are similar: $w$ is essentially unchanged on passing $\beta_{c1}$, and increases by only about $10\%$ over the entire range $\beta_{c1} \rightarrow \beta_{c2}$. This provides evidence that the staircase-like steps, observed in $N_D$ for moderate and large $|U|$, can be attributed to pair breaking as opposed to pair loosening.

\section{Phenomenological model}
\label{sec:phenomenological-model}

Having established the basic behavior of the ground state, we will next present a phenomenological model which integrates our findings from previous sections to approximate the nature of the ground state at different $\beta$. This will allow us to derive approximate expressions for the thresholds $\beta_{c1}$ and $\beta_{c2}$.

In Section~\ref{sec:exact-diagonalization} it was shown that changes in the ground-state bound pair number can be described in terms of crossings between the ground state, and eigenstate manifolds which correspond to specific $N_D$ values (and roughly correspond to specific integer numbers $M = 0,1, \ldots$ of bound pairs). Therefore, a way to estimate the ground state's pair number is to estimate the lowest energy of each manifold at given $\beta$ and determine which manifold has the lowest overall energy. As we do not have analytical Bethe ansatz solutions for the eigenstates at $\beta \ne 0$, we instead estimate these eigenenergies from a simple phenomenological model.

To start we consider the lowest-energy level of a manifold corresponding to a particular number of bound pairs $M$ (assumed based on the value of $N_D$). Let us estimate the energy of this lowest level. We assume that the energy of each bound pair is given by the two-body bound-state energy of the one-dimensional Hubbard model~\cite{KORNILOVITCH2024169574}, $\Delta(U,P) = -\sqrt{U^2 + 16\cos^2(P/2)}$, where $P$ is the pair center-of-mass momentum. For simplicity, we set $P=0$ for all pairs, so that $\Delta(U) = -\sqrt{U^2+16}$. The pair energy is assumed to be independent of $\beta$, which is justified in the high-$|U|$ limit where each pair is nearly a doublon and feels a negligible net external potential (for moderate $|U|$ this is an approximation). The remaining unpaired fermions are assumed to occupy the lowest single-particle orbitals of their respective spin bands, with $\beta$-dependent energies $\mathcal{E}_{1}, \ldots, \mathcal{E}_{N_\uparrow-M}$ and $\mathcal{E}_{1}, \ldots, \mathcal{E}_{N_\downarrow-M}$. The lowest energy of the manifold with $M$ pairs is then estimated as
\begin{equation}
    \label{eq:model-energy-estimate}
    E_M(\beta) = M \left(-\sqrt{U^2+16}\right) + \sum_{\sigma=\uparrow,\downarrow} \sum_{n=1}^{N_\sigma - M} \mathcal{E}_{n}(\beta).
\end{equation}
Therefore, at each $\beta$ the ground-state energy can be estimated as $\min_M[E_M(\beta)]$, with the corresponding $M$ giving the ground-state pair number.

\begin{figure}[t]
    \centering
    \includegraphics[width=0.95\linewidth]{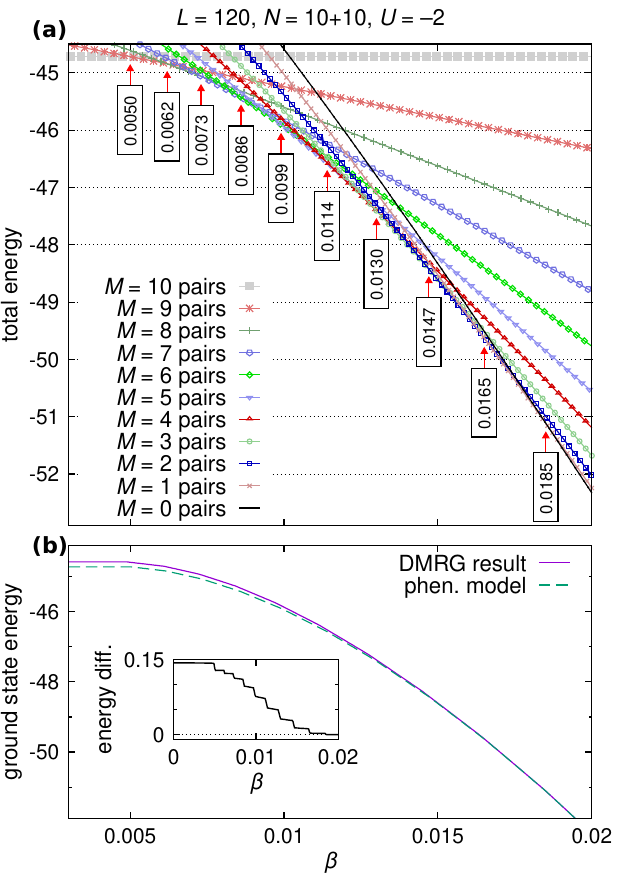}
    \caption{(a) The estimated energy $E_M(\beta)$ (from Eq.~\eqref{eq:model-energy-estimate}) for all possible pair numbers $M = 0, \ldots, 10$, for a system with $L = 120, U = -2, N = 10 + 10$. Red arrows, and their accompanying labels, indicate the values of $\beta$ at which adjacent lowest-energy curves cross. (b) Comparison between the DMRG ground-state energy and the phenomenological model estimate, obtained as the minimum of $E_M(\beta)$ over $M$. Inset: The difference between the two energy predictions.}
    \label{fig:energy-switches-in-phenomenological-model}
\end{figure}

In Fig.~\ref{fig:energy-switches-in-phenomenological-model}(a), we plot the estimated energies $E_M(\beta)$ for each possible $M$ in a system with $L=120, N=10+10, U=-2$. The single-particle energies $\mathcal{E}_{n}(\beta)$ are found numerically, by exact diagonalization of the single-particle Hamiltonian at each $\beta$. The curves $E_M(\beta)$ show a series of crossings, with the lowest energy $\min_{M}\left[E_M(\beta)\right]$ successively assumed by decreasing values of $M$ (each crossing is indicated in Fig.~\ref{fig:energy-switches-in-phenomenological-model} by a red arrow). Therefore, the model predicts one-by-one decreases in pair number with increasing $\beta$, in accordance with the DMRG results. In Fig.~\ref{fig:energy-switches-in-phenomenological-model}(b), we compare the resulting prediction for the ground-state energy with the DMRG result. The inset shows the difference between the two energies, which decreases in a stepwise manner as $\beta$ increases, so the phenomenological estimate becomes more accurate after each pair-breaking event. This indicates that the assumption of a fixed energy $-\sqrt{U^2+16}$ per pair is only approximate, with the energy error being largest when many pairs are present. Nevertheless, the absolute energy curves in Fig.~\ref{fig:energy-switches-in-phenomenological-model}(b) are qualitatively similar, confirming that the model correctly captures the overall trends.

\begin{figure}[t]
    \centering
    \includegraphics[width=0.95\linewidth]{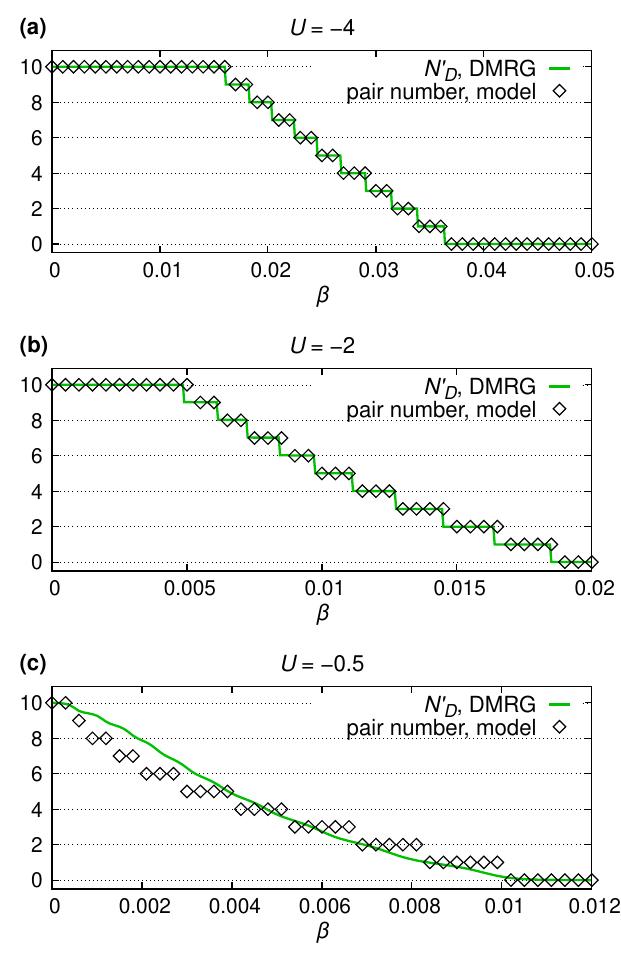}
    \caption{Rescaled doublon number $N'_D$ from DMRG (green line), and predicted pair number $M$ from the phenomenological model, given by $M = \mathrm{argmin}_M E_M(\beta)$ (diamond symbols), for three values of $U$ with $L=120, N=10+10$.}
    \label{fig:doublon-number-comparison-between-dmrg-and-phen-model}
\end{figure}

This model can be used to predict the critical $\beta$ corresponding to each $M \to M-1$ ground-state transition. In Fig.~\ref{fig:doublon-number-comparison-between-dmrg-and-phen-model}, we compare the rescaled doublon number $N'_D$ from DMRG with the predicted pair number $M$ from the model. In panels (a) and (b) ($U = -4$ and $U = -2$, respectively), the location of each transition is very accurately estimated by the model. Panel (c) ($U = -0.5$) shows that at small $|U|$ the model is only approximate, since it always predicts sharp transitions and does not capture the effects that smooth out the steps in $N'_D$.

The two threshold values $\beta_{c1}$ and $\beta_{c2}$ can be estimated by considering the energy cost of converting the first and the last pair, respectively, into two separated fermions. This gives the predictions
\begin{align}
2 \mathcal{E}_1(\beta_{c1}) &= \Delta(U), \label{eq:numerical-estimate-for-betac1} \\
2 \mathcal{E}_{N/2}(\beta_{c2}) &= \Delta(U), \label{eq:numerical-estimate-for-betac2}
\end{align}
where we have assumed $N_\uparrow = N_\downarrow = N/2$. We omit the spin indices on $\mathcal{E}$, since the energies are spin-independent. The above formulas can be used directly, by computing $\mathcal{E}_n(\beta)$ numerically (via diagonalization of the single-body Hamiltonian) and substituting into Eqs.~\eqref{eq:numerical-estimate-for-betac1} and~\eqref{eq:numerical-estimate-for-betac2}. To obtain closed-form analytical approximations to $\beta_{c1}$ and $\beta_{c2}$, one can instead use analytical estimates for $\mathcal{E}_{1}(\beta)$ and $\mathcal{E}_{N/2}(\beta)$.

\subsection{Analytical approximations of $\beta_{c1}, \beta_{c2}$}

To estimate $\beta_{c1}$, we use the approximation for $\mathcal{E}_1$ given in Eq.~\eqref{eq:lowest-energy-approximation}. Following the reasoning detailed in Appendix~\ref{sec:phenomenological-model-estimate-betac1}, we find
\begin{align}
\label{eq:beta-c1-expression-to-second-order}
\beta_{c1} (L+1) &\approx \left(\sqrt{U^2+16} -4 \right) \\
&+ \left(\sqrt{U^2+16} -4\right)^{2/3} \frac{2\cdot 2.270}{(L+1)^{2/3}} \nonumber \\
&+ O\left(L^{-4/3}\right). \nonumber
\end{align}
The $O\left(L^{-2/3}\right)$ and higher-order terms are finite-size corrections, originating from the influence of the lattice edge on the lowest orbital energy $\mathcal{E}_1$. Finite-size corrections are important for $\beta_{c1}$ because the first broken pair decomposes into fermions localized near lattice edges. Overall, Eq.~\eqref{eq:beta-c1-expression-to-second-order} predicts that $\beta_{c1} L$ is independent of the particle number, and, to leading order, becomes linear in $U$ as $|U| \rightarrow \infty$. These predictions are generally consistent with our earlier observations in Sec.~\ref{sec:dependency-of-critical-beta-on-parameters}.

To estimate $\beta_{c2}$, we assume that the energy $\mathcal{E}_{N/2}$ lies in the middle part of the single-particle spectrum, corresponding to the Wannier--Stark regime, so that it can be approximated by the on-site potential:
\begin{equation}
    \mathcal{E}_{N/2} \approx \beta \left( \frac{N}{2} - \frac{L+1}{2} \right). \label{eq:wannier-stark-approximation-of-energy}
\end{equation}
This approximation is particularly well justified for $n\approx 1$. Substituting into Eq.~\eqref{eq:numerical-estimate-for-betac2} gives
\begin{equation}
    2 \beta_{c2} \left(\frac{N}{2} - \frac{L+1}{2}\right) \approx -\sqrt{U^2+16},
\end{equation}
which yields
\begin{equation}
\label{eq:beta-c2-estimate}
    \beta_{c2} L \approx \frac{\sqrt{U^2+16}}{1-n+\frac{1}{L}}.
\end{equation}

This result predicts that $\beta_{c2} L$ becomes linear in $U$ for large $|U|$, increases monotonically with $n$, and is independent of $L$ in the large-$L$ limit. These predictions generally agree with the dependence on $U, n, L$ observed in the DMRG calculations (Fig.~\ref{fig:critical-beta-values}). The $1/L$ term represents a finite-size correction, which is less important than in the case of $\beta_{c1} L$, because the last broken pair is assumed to decompose into fermions occupying bulk-localized Wannier--Stark orbitals.

\begin{figure}[t]
    \centering
    \includegraphics[width=0.95\linewidth]{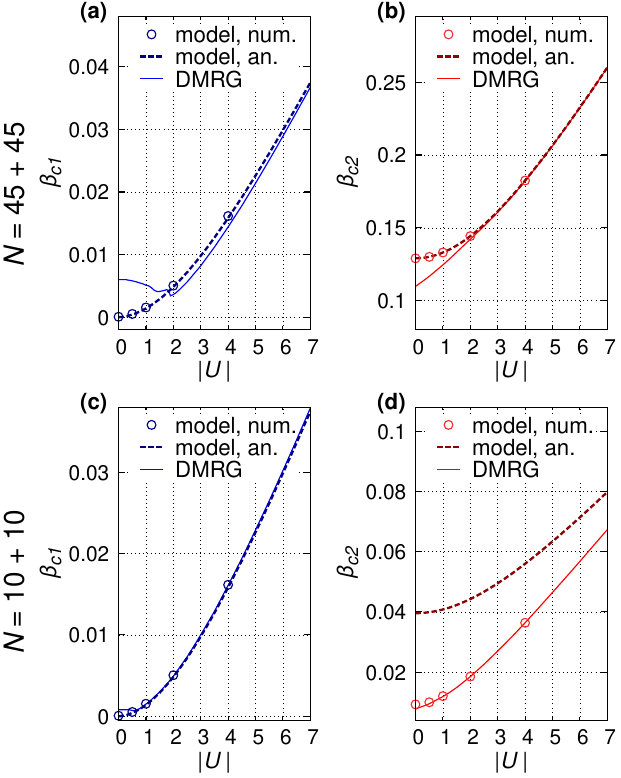}
    \caption{(a) Threshold value $\beta_{c1}$ as a function of $|U|$ for $L=120$, $N=45+45$, obtained in three ways: from DMRG (continuous line), from the analytical estimate in Eq.~\eqref{eq:beta-c1-expression-to-second-order} (dashed line), and from substituting numerically determined $\mathcal{E}_n(\beta)$ into Eq.~\eqref{eq:numerical-estimate-for-betac1} (circles). (b) Threshold values $\beta_{c2}$ obtained from DMRG (continuous line), from the analytical estimate in Eq.~\eqref{eq:beta-c2-estimate} (dashed line), and from substituting numerically determined $\mathcal{E}_n(\beta)$ into Eq.~\eqref{eq:numerical-estimate-for-betac2} (circles).  (c,d) Same for $N=10+10$.}
    \label{fig:comparison_dmrg_vs_phenmodel_thresholds}
\end{figure}

In Fig.~\ref{fig:comparison_dmrg_vs_phenmodel_thresholds}, we compare the phenomenological model estimates for $\beta_{c1}$ and $\beta_{c2}$ with the DMRG results. We consider: direct numerical estimates, which we have obtained by computing $\mathcal{E}_n(\beta)$ numerically and substituting into Eqs.~\eqref{eq:numerical-estimate-for-betac1} and~\eqref{eq:numerical-estimate-for-betac2}; and the closed-form analytical estimates given respectively by Eqs.~\eqref{eq:beta-c1-expression-to-second-order} (to subleading order) and~\eqref{eq:beta-c2-estimate}.

For $N=45 + 45$ [Fig.~\ref{fig:comparison_dmrg_vs_phenmodel_thresholds}(a,b)], the model estimates deviate from the DMRG results at small $|U|$, but at larger $|U|$, both $\beta_{c1}$ and $\beta_{c2}$ are predicted accurately. This confirms that the accuracy of the phenomenological model is tied to the well-defined pair number that emerges at stronger interactions, consistent with the results in Fig.~\ref{fig:doublon-number-comparison-between-dmrg-and-phen-model} where the model predictions are accurate only at strong $|U|$. The analytical estimates agree closely with the numerical ones, indicating that the analytical approximations used for $\mathcal{E}_n$ are accurate in this regime.

The situation is different for $N=10+10$ [Fig.~\ref{fig:comparison_dmrg_vs_phenmodel_thresholds}(c,d)]. For this smaller population, the DMRG result for $\beta_{c1}$ is still reproduced correctly by the model, apart from deviations at very small $|U|$. The numerical estimate for $\beta_{c2}$ is also highly accurate. However, the analytical estimate deviates from the numerical estimate and significantly overestimates $\beta_{c2}$, especially at low $|U|$. This indicates that the Eq.~\eqref{eq:beta-c2-estimate} estimate is too crude for small particle numbers, where the single-particle energies $\mathcal{E}_{N/2}$ are not near the middle of the band and are not well approximated by Eq.~\eqref{eq:wannier-stark-approximation-of-energy}.

As a sanity check, it is worth checking if the estimates for $\beta_{c1}$ and $\beta_{c2}$ behave correctly in the $N = 1+1$ case, where we expect $\beta_{c1} = \beta_{c2}$ by definition. Assuming ${L \to +\infty}$, from Eqs.~\eqref{eq:beta-c1-expression-to-second-order} and ~\eqref{eq:beta-c2-estimate} we find
\begin{equation}
\label{eq:estimate-betac1c2ratio-phenmodel}
    \frac{\beta_{c2}}{\beta_{c1}} \bigg|^\mathrm{phen.model}_{N=2} \approx 1 + \frac{4}{\sqrt{U^2+16} - 4},
\end{equation}
which is not unity except in the $|U| \to +\infty$ limit. This further shows that the $\beta_{c2} L$ estimate is poor at small $N$.

\section{Local-density approximation analysis}
\label{sec:lda}

The local-density approximation (LDA) is a common approach for analyzing the equilibrium states of systems in non-uniform potentials. This method approximates the ground state of a non-uniform lattice by relating each site to a separate uniform system, governed by a suitably defined local chemical potential. In this section, we assess the applicability of LDA to our model by comparing its predictions with DMRG results.

\subsection{Description of the LDA approach}
\label{sec:lda-description}

Within the LDA, we treat the inhomogeneous system by first considering a reference system: an infinite, translation-invariant Hubbard model without the external potential, described in the grand-canonical ensemble by
\begin{align}
  \label{eq:grand-canonical-energy-functional}
  \hat K(\mu,h) = &-t \sum_{j} \sum_{\sigma=\uparrow,\downarrow} \left[ \hat{c}^\dagger_{j,\sigma} \hat{c}_{j+1,\sigma} + \mathrm{H.c.} \right] \nonumber \\
  &+ U \sum_{j} \hat{n}_{j,\uparrow} \hat{n}_{j,\downarrow} - \mu_\uparrow \hat{N}_\uparrow - \mu_\downarrow \hat{N}_\downarrow,
\end{align}
where $\mu_\sigma$ is a constant chemical potential felt by spin component $\sigma$. It is convenient to also define the combined quantities of the average chemical potential ${\mu = (\mu_\uparrow+\mu_\downarrow)/2}$ and the effective Zeeman field ${h = (\mu_\uparrow-\mu_\downarrow)/2}$. For this translation-invariant reference system, the ground-state equation of state---\emph{i.e.}, the functions $n_\sigma(\mu,h;U)$ and $n_D(\mu,h;U)$---can be determined by analytical or numerical means.

In the presence of the spin-dependent linear potential $V_{j,\sigma}$, the local chemical potentials vary on each site $j$ as
\begin{equation}
  \mu_\sigma(j) = \mu_\sigma - V_{j,\sigma},
\end{equation}
so that $\mu(j)=\mu$ is constant while $h(j) = h - \beta\left(j-j_0\right)$ varies linearly across the lattice. The LDA then approximates the density on each site by sampling the homogeneous equation of state at these local parameters:
\begin{align}
  n^{\mathrm{LDA}}_{j,\sigma} &= n_\sigma \left(\mu(j), h(j);U\right), \nonumber \\
  n^{\mathrm{LDA}}_{j,D} &= n_D\left(\mu(j), h(j);U\right).
  \label{eq:lda-equation-of-state}
\end{align}

The LDA has certain limitations. First, it assumes that the external potential varies slowly enough to be regarded as locally constant, requiring that it not change appreciably over the spatial extent of the relevant local correlations, such as the size of a bound pair. Therefore, the LDA is generally not applicable to arbitrarily high $\beta$. Moreover, due to its local nature, the LDA cannot predict the full ground-state wave function or observables that depend on long-range correlations, such as momentum distributions; it can, however, approximate the density profiles. Since the transitions in our ground state are visible already at the density distribution level, this capability is sufficient for our purposes.

To determine the equation of state [Eq.~\eqref{eq:lda-equation-of-state}], we use the known phase diagram of the attractive Fermi-Hubbard model expressed in the $\mu$ and $h$ coordinates~\cite{HeidrichMeisner2010,essler2005}, which yields the equilibrium densities $n_\sigma(\mu,h;U)$ for each pair of parameter values $(\mu,h)$. Within the LDA, the ground-state density of the inhomogeneous system corresponds to a trajectory which crosses the phase diagram from point $(\mu,h(j=1))$ to $(\mu,h(j=L))$, where the global values $\mu, h$ are chosen so as to recover the desired total populations $N_\uparrow, N_\downarrow$ upon summing over all sites. The resulting density distribution reflects the sequence of phase regions traversed by this trajectory.

To obtain LDA results for the system with given $\beta, N_\uparrow, N_\downarrow$ and $U$, we first construct the phase diagram at the given $U$ as a dense grid of values $n_\sigma(\mu,h;U)$ and $n_D(\mu,h;U)$; see Appendix~\ref{sec:phase-diagram} for details on the used procedure. We then determine the global values of $\mu$ and $h$ by scanning over candidate values until the total density along the trajectory $\left(\mu, h+\beta\frac{L-1}{2}\right) \to \left(\mu, h-\beta\frac{L-1}{2}\right)$ matches the desired fixed particle numbers for both spin components. (For $N_\uparrow = N_\downarrow$, the global Zeeman field always vanishes, $h=0$.) Once $\mu$ and $h$ are determined, the density at site~$j$ is estimated via Eq.~\eqref{eq:lda-equation-of-state}.

\subsection{LDA results for the density profile}
\label{sec:lda-results}

\begin{figure}[t]
    \centering
    \includegraphics[width=0.95\linewidth]{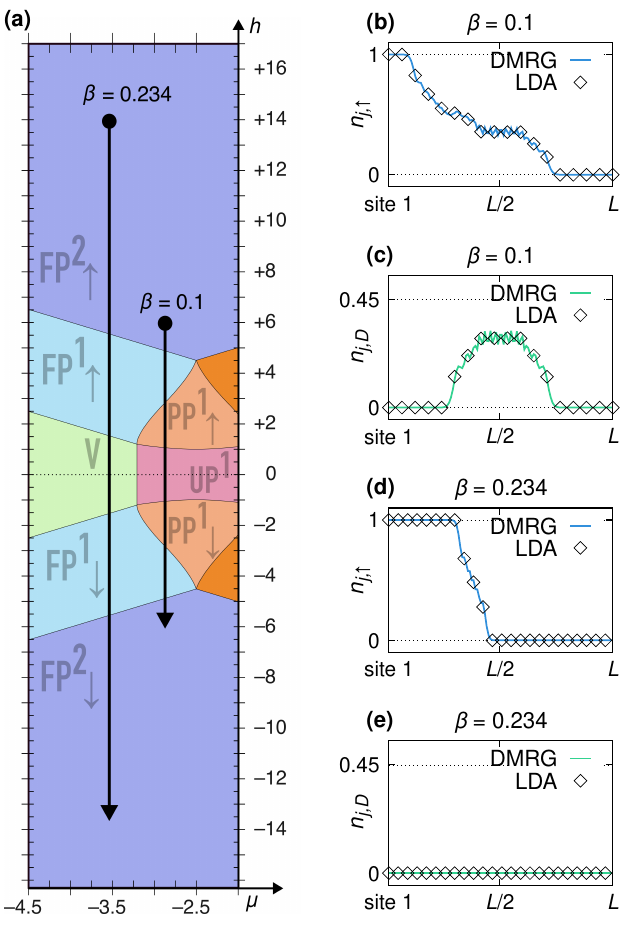}
    \caption{Example of predicting the density profile via the local-density approximation. (a) Phase diagram of the grand-canonical Hubbard model, in the $\mu$-$h$ plane, for $U=-5$ (see Appendix~\ref{sec:phase-diagram} for explanation of phase labels). The two trajectories across the phase diagram correspond to lattice systems with the parameters $U = -5$, $N = 45+45$, $L=120$, and two different gradients: $\beta = 0.1$, $\beta = 0.234$. The two trajectories are defined by the following coordinates: $\mu = -2.867$, $h(1) = 5.95$, $h(L) = -5.95$, and: $\mu = -3.535$, $h(1) = 13.923$, $h(L) = -13.923$, respectively. Each trajectory crosses several phase regions. (b,c) Density profiles ($n_{j,\uparrow}$ and $n_{j,D}$) for the $\beta=0.1$ system, predicted by DMRG (lines), and by the LDA per the trajectory in (a) (symbols). (d,e) Same, for the $\beta=0.234$ system.}
    \label{fig:trajectory-example}
\end{figure}

\begin{figure}[t]
    \centering
    \includegraphics[width=0.95\linewidth]{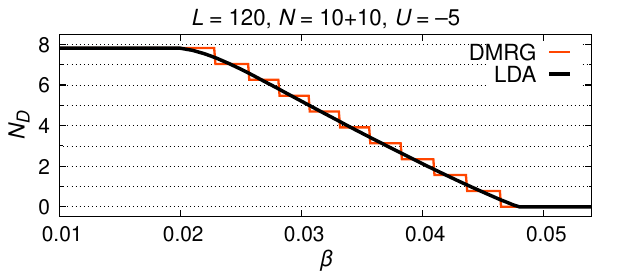}
    \caption{Doublon number $N_D$ as a function of $\beta$ for $U = -5$, $L=120$, $N=10+10$, as estimated by DMRG (thin line) and LDA (thick line).}
    \label{fig:doublon-number-lda-dmrg-comparison}
\end{figure}

As an example, Fig.~\ref{fig:trajectory-example}(a) shows two LDA trajectories, overlaid on the phase diagram (obtained as described in Appendix~\ref{sec:phase-diagram}). Both trajectories correspond to systems with $U = -5$, $N = 45+45$, $L=120$, but at two different gradient strengths: $\beta = 0.100$ and $\beta = 0.234$. For these parameters, $\beta_{c1} \approx 0.02$ and $\beta_{c2} \approx 0.21$, so the two trajectories provide examples of the phase-separated regime ($\beta_{c1} \le \beta < \beta_{c2}$) and the spin-separated regime ($\beta \ge \beta_{c2}$), respectively. Each trajectory runs from $(\mu, h(1))$ to $(\mu, h(L))$, with the lattice-wide chemical potential determined by scanning over values of $\mu$ until the total LDA density matched $N_\uparrow = N_\downarrow = 45$.

The trajectory for $\beta=0.100$ crosses several phase regions: the point corresponding to site 1 lies in the fully-polarized $\mathrm{FP}_\uparrow$ phase, and for subsequent sites the trajectory passes through the partially polarized ($\mathrm{PP}_\uparrow$) and unpolarized ($\mathrm{UP}$) phases before crossing into $\mathrm{PP}_\downarrow$ and reaching $\mathrm{FP}_\downarrow$ at site $L$. This sequence matches the expected picture of polarized opposite-spin wings surrounding an unpolarized core, with partially polarized buffer regions in between. The accompanying density comparisons [Fig.~\ref{fig:trajectory-example}(b,c)] for $n_{j,\uparrow}$ and $n_{j,D}$ show that the LDA densities agree closely with the DMRG calculations, except for missing finite-size features such as the Friedel oscillations of density in the paired core.

Similarly, the $\beta=0.234$ trajectory crosses the fully-polarized ($\mathrm{FP}$) and vacuum ($\mathrm{V}$) regions, in a sequence that corresponds to a spin-separated system with spin components congregated on opposite edges. The density comparisons are shown in Fig.~\ref{fig:trajectory-example}(d,e) and show that the LDA and DMRG results also agree closely in this case.

We also examine whether $N_D$ can be estimated via the LDA. In Fig.~\ref{fig:doublon-number-lda-dmrg-comparison}, we compare $N^\mathrm{LDA}_D = \sum_{j=1}^L n_{j,D}^\mathrm{LDA}$ with the DMRG result for $N_D$ as a function of $\beta$, in a system with $N = 10+10$ and $U = -5$. The LDA reproduces the DMRG values of $N_D(\beta)$ reasonably well at each $\beta$, and correctly predicts the overall trend: a monotonic decrease that begins and ends abruptly at values of $\beta$ close to the DMRG-derived $\beta_{c1}$ and $\beta_{c2}$. However, the LDA fails to capture the discrete stepwise structure, instead predicting a continuous, approximately linear decrease.

In summary, the LDA density distributions accurately recreate the DMRG results. The LDA picture also provides an alternate way to understand and predict the sequence of distinct polarization regions across the lattice, by analyzing the trajectory the system traces through the phase diagram. While the behavior of $N_D$ can be also approximately recovered by the LDA, it misses important features, notably the discrete steps that reflect one-by-one pair breaking.

\subsection{LDA-based estimate for the $\beta_{c2}$ threshold}
\label{sec:lda-betac2-estimate}

The LDA can also be used to derive an improved analytical estimate for $\beta_{c2}$ that, unlike the Wannier--Stark-based approximation in Eq.~\eqref{eq:beta-c2-estimate}, works well for small populations. It is based on the fact that, for LDA trajectories corresponding to the $\beta > \beta_{c2}$ spin-separated ground states (\emph{i.e.}, trajectories passing only through the regions $\mathrm{V}$, $\mathrm{FP}^1_\sigma$, and $\mathrm{FP}^2_\sigma$), one can establish exact relationships between $\mu$, $\beta$, $U$, and $N_\sigma$. The corresponding calculations are carried out in Appendix~\ref{sec:lda-betac2-estimate}. The results are:

\begin{itemize}
\item For small population, \emph{i.e.}, when $\beta_{c2}$ is small enough that $N_\uparrow = N_\downarrow < \frac{2}{\beta_{c2}}$, the estimate for $\beta_{c2}$ is given by a root of the quadratic equation
\begin{align}
\label{eq:square-equation-for-betac2}
    \beta_{c2}^2 \cdot  &\frac{(L-1)^2}{4}   \\
    + \beta_{c2} \cdot &\left[ (L-1)\left(\pi-\frac{1}{2}\sqrt{U^2 + 16}\right)  - 4 \pi N_\uparrow\right] \nonumber \\
    + &\left[ 12 + \frac{U^2}{4} - \pi\sqrt{U^2+16} \right] = 0. \nonumber
\end{align}

\item For large population, $N_\uparrow = N_\downarrow \ge \frac{2}{\beta_{c2}}$, the estimate is 
\begin{equation}
    \beta_{c2} L = \frac{\sqrt{U^2+16}}{1-n-\frac{1}{L}},
\end{equation}
which is nearly identical to the phenomenological model estimate in Eq.~\eqref{eq:beta-c2-estimate}, differing only in the sign of a $1/L$ term. In the large-$L$ limit, the two expressions are equivalent.
\end{itemize}

\begin{figure}[t]
    \centering
    \includegraphics[width=0.95\linewidth]{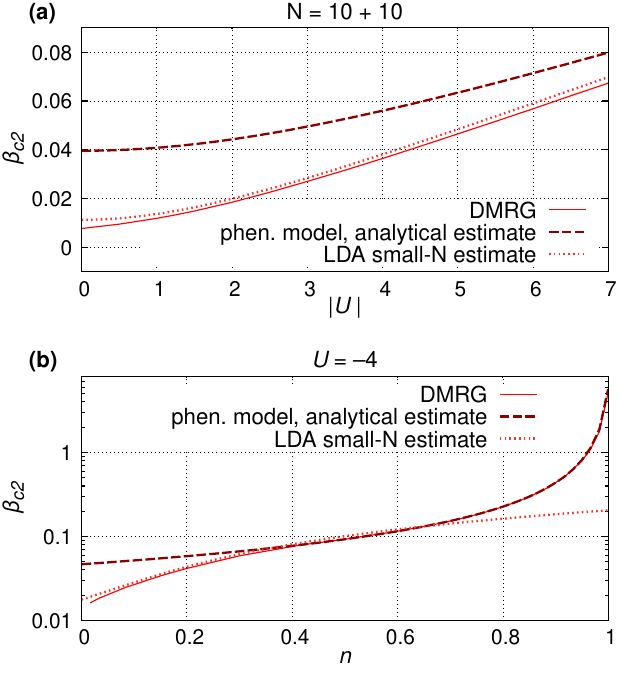}
    \caption{(a) Threshold value $\beta_{c2}$ as a function of $|U|$ for $L=120$, $N=10+10$. Three estimates are shown: the DMRG result (continuous line), the phenomenological analytical estimate from Eq.~\eqref{eq:beta-c2-estimate} (dashed line), and the LDA-based small-population estimate (root of Eq.~\eqref{eq:square-equation-for-betac2}; dotted line). (b) Threshold value $\beta_{c2}$ as a function of $n$ for $L=120$, $U = -4$. Note the logarithmic scale.}
    \label{fig:critical-beta_refined-LDA-and-phenmodel-vs-dmrg}
\end{figure}

In Fig.~\ref{fig:critical-beta_refined-LDA-and-phenmodel-vs-dmrg}, we compare the small-population LDA estimate with the DMRG-derived values and with the Wannier--Stark-based analytical prediction of Eq.~\eqref{eq:beta-c2-estimate}. Figure~\ref{fig:critical-beta_refined-LDA-and-phenmodel-vs-dmrg}(a) shows $\beta_{c2}$ as a function of $|U|$ for $N=10+10$. The LDA estimate closely tracks the DMRG prediction, significantly improving on the analytical approximation. Fig.~\ref{fig:critical-beta_refined-LDA-and-phenmodel-vs-dmrg}(b) shows the result for $U = -4$ and for varying populations $n$. As expected, at low $n$ the LDA-based estimate is a close approximation to the DMRG result, while the Wannier--Stark approximation fails; at higher $n$, the small-population LDA approximation breaks down, while the Wannier-Stark approximation becomes accurate.

Once again, it is worth checking if this estimate predicts $\beta_{c1}=\beta_{c2}$ in the $N = 1+1$ limit. Assuming $L \to +\infty$, from Eqs.~\eqref{eq:beta-c1-expression-to-second-order} and ~\eqref{eq:square-equation-for-betac2} we find
\begin{align}
\label{eq:estimate-betac1c2ratio-lda}
    \frac{\beta_{c2}}{\beta_{c1}} \bigg|^\text{LDA estimate}_{N=2} &\approx \frac{\sqrt{U^2+16} - \left(2\pi - 2\sqrt{\pi^2-8}\right)}{\sqrt{U^2+16} - 4} \\
    &\approx \frac{\sqrt{U^2+16} - 3.55}{\sqrt{U^2+16} - 4}, \nonumber
\end{align}
which, while not exactly unity, converges to unity with increasing $|U|$ and, for every $|U|$, lies much closer to unity than the phenomenological model estimate in Eq.~\eqref{eq:estimate-betac1c2ratio-phenmodel}.

In conclusion, by combining the local-density approximation with the phenomenological model, one can obtain a set of closed-form expressions that predict $\beta_{c1}$ and $\beta_{c2}$ as functions of $|U|$, $n$, and $L$. Together, these expressions cover the full range of $|U|$ and $n$.

\section{Ground-state properties under added harmonic confinement}
\label{sec:harmonic}

In cold-atom experiments with one-dimensional lattices, an additional harmonic confinement is typically present, arising from the combination of the trapping potential and the Gaussian intensity profile of the lattice beams~\cite{bloch2005ultracold}. To assess the robustness of our results, in this section we add a spin-independent harmonic term to the external potential and examine whether the characteristic staircase-like behavior of $N_D$ persists.

\begin{figure}[t]
    \centering
    \includegraphics[width=0.94\linewidth]{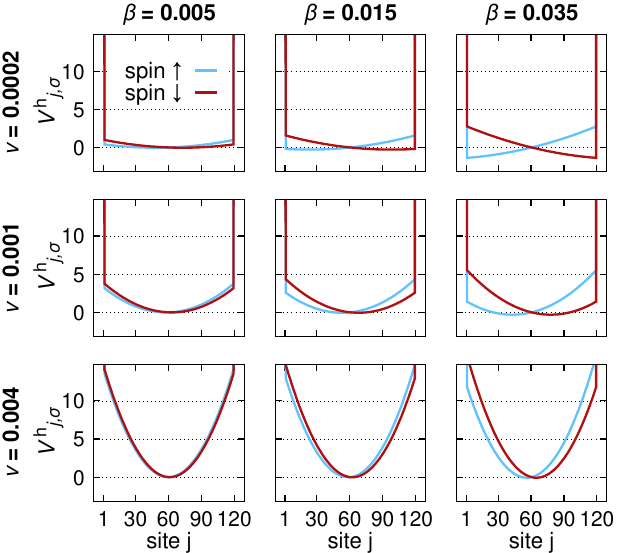}
    \caption{Modified external potentials $V^\mathrm{h}_{j,\sigma}$ [Eq.~\eqref{eq:potential-with-harmonic-term}] for several values of the harmonic confinement strength $v$ and the linear potential gradient $\beta$. Both $V^\mathrm{h}_{j,\uparrow}$ (blue) and $V^\mathrm{h}_{j,\downarrow}$ (red) are shown.}
    \label{fig:external-potential-shape-with-harmonic-trap}
\end{figure}

We consider a modified external potential
\begin{equation}
    \label{eq:potential-with-harmonic-term}
    V^\mathrm{h}_{j,\sigma} = V_{j,\sigma} + v\, (j-j_0)^2,
\end{equation}
where $v$ (in units of $t$) is the harmonic confinement strength. This potential is equivalent to a shifted harmonic trap:
\begin{equation}
    \label{eq:potential-with-harmonic-term-altform}
    V^\mathrm{h}_{j,\sigma} = v\, \left(j-j_0 \pm \frac{\beta}{2v}\right)^2 + \text{const},
\end{equation}
where $\sigma=\uparrow,\downarrow$ corresponds to $\pm = +,-$ respectively. Note that we retain the open boundary conditions with a hard-wall cutoff at sites $1$ and $L$, so the potential seen by the fermions is not precisely harmonic. However, for sufficiently strong harmonic confinement, the particles are confined in the bulk away from the lattice ends, and the physics becomes insensitive to the boundary conditions. Examples of the resulting potentials $V^\mathrm{h}_{j,\sigma}$ are shown in Fig.~\ref{fig:external-potential-shape-with-harmonic-trap}. 

\begin{figure}[t]
    \centering
    \includegraphics[width=0.94\linewidth]{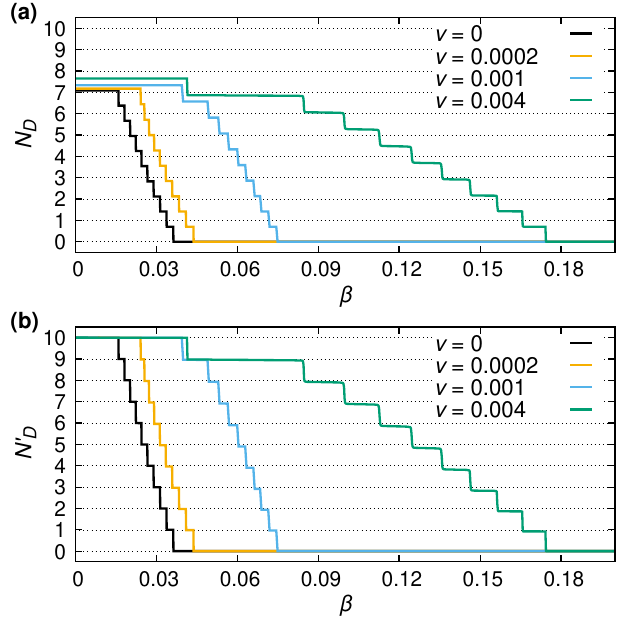}
    \caption{(a) $N_D$ and (b) rescaled $N'_D$ as a function of $\beta$, for a system with $U = -4$, $L=120$, $N=10+10$ and different strengths $v$ of harmonic confinement.}
    \label{fig:doublon-number-under-harmonic-confinement}
\end{figure}

In Fig.~\ref{fig:doublon-number-under-harmonic-confinement}, we examine the numerically determined doublon number $N_D$ and its rescaled counterpart $N'_D$ under harmonic confinement for $L=120$, $N=10+10$, $U=-4$, and several values of $v$. The staircase-like structure of $N'_D$, with steps close to integer numbers, remains intact under harmonic confinement. The pair-breaking thresholds $\beta_{c1}$ and $\beta_{c2}$ shift upward with increasing harmonic confinement, which can be understood intuitively: the harmonic potentials push both spin components toward the center of the lattice, so a stronger gradient $\beta$ is required to fully separate them.

Overall, this initial check suggests that our conclusions are not qualitatively altered by the presence of harmonic confinement. More broadly, it hints that the hard-wall boundary conditions assumed in this work may not be essential, and that similar pair-breaking behavior could arise in a wider class of setups where the two spin components experience external potentials with spatially separated minima. A systematic study of this possibility is left for future work.

\section{Conclusion}
\label{sec:conclusion}

We have investigated the ground state of a one-dimensional, open-boundary attractive Fermi--Hubbard model subject to an external linear potential with opposite gradients for the two spin components. By varying the gradient strength $\beta$ together with the interaction $U$, system size $L$, and filling $n$, we have mapped out how the competition between spin-separating forces and onsite attraction shapes the ground state.

As $\beta$ increases, the system evolves from a fully paired state to a fully spin-separated configuration via successive, one-by-one breaking of bound pairs at specific critical values of $\beta$. Three $\beta$ regimes can be distinguished, with qualitatively different ground state behavior. Below a first threshold $\beta_{c1}$, the densities of both spin components are nearly identical, and the doublon number is insensitive to increase of $\beta$. Above a second threshold $\beta_{c2}$, all pairs are broken and the ground-state density is entirely spin-separated, with vanishing doublon number. For $\beta$ tuned between $\beta_{c1}$ and $\beta_{c2}$, the doublon number decreases in discrete steps representing breaking of successive pairs, and the ground-state density is phase-separated, with a shrinking paired core flanked by growing polarized wings. The number of remaining bound pairs can be quantified in two ways: from the decrease in the doublon number relative to its value at $\beta=0$, or from the total excess density of the local majority component across the lattice. The signatures of pair breaking are most pronounced at strong interactions $|U|$, but they remain discernible for weaker attractions. We provide a simple theoretical picture of the transitions between successive bound-pair numbers, study them in detail via numerical calculations and the local-density approximation, and derive rough estimates for the critical gradient values.

Our results demonstrate that spin-dependent linear potentials offer a precise means for controlling the number of paired fermions in the lattice: by tuning $\beta$, the number of bound pairs can be stabilized at distinct integer values, each of which is maintained over a finite range of $\beta$. The predicted relationships between the critical gradients and the system parameters provide concrete experimental observables for future research.

There are several possible directions for future work. First, one could analyze pairing correlations and the associated Cooper pair-like signatures in the presence of a finite gradient, in order to clarify how superconducting correlations change as the system approaches the spin-separated regime. We provide a starting point for this analysis in Appendix~\ref{sec:fflo-signatures}, where we present FFLO-like signatures for one particular parameter set. Second, studying the finite-temperature behavior of the transitions identified here would help assess the experimental feasibility of observing the phase-separated density regime. Third, dynamical protocols such as slow ramps or sudden quenches of the gradient could be used to track pair breaking and spin separation in real time.

In this work we have focused on balanced systems ($N_\uparrow = N_\downarrow$), but the extension to unbalanced populations is straightforward. A finite spin imbalance would break the spatial and spin symmetries of the ground states. Another natural generalization is to introduce unequal masses or hopping amplitudes for the two spin species, which is experimentally feasible in cold-atom mixtures of different atomic species or hyperfine states. Such extensions would further enrich the interplay between spin-dependent forces and pairing. Finally, while we have verified that the staircase-like pair-breaking structure persists under the addition of spin-independent harmonic confinement, a comprehensive study of the system properties under such an additional potential remains an open direction.

The datasets and processed data files used to generate the plots and analyses in this study are openly available in Zenodo~\cite{zenodo}.

\begin{acknowledgements}
This work was supported by the Okinawa Institute of Science and Technology Graduate University (OIST) and utilized the computing resources of the Scientific
Computing and Data Analysis section of Core Facilities at OIST. It was also supported by the Japan Science and Technology Agency (JST) Grant No. JPMJPF2221.
\end{acknowledgements}

\appendix

\section{Useful many-body Hamiltonian symmetries}
\label{sec:many-body-hamiltonian-symmetries}

The many-body Hamiltonian $\hat{H}$ in Eq.~\eqref{eq:many-body-hamiltonian} has several useful symmetries.

The first thing to note is that applying either a spin-flip $(\uparrow,\downarrow) \rightarrow (\downarrow,\uparrow)$ or a spatial reflection ($j \rightarrow L+1-j$) to all field operators in $\hat{H}$ is equivalent to inverting the external potential ($\beta \rightarrow -\beta$). By using this fact, the ground state obtained for given $(N_\uparrow,N_\downarrow,\beta)$ can be easily extrapolated to that at inverted ${\beta \to -\beta}$ or exchanged populations ${N_\uparrow \leftrightarrow N_\downarrow}$.

The many-body Hamiltonian also exhibits particle-hole duality: a direct correspondence exists between the eigenstates in the $(N_\uparrow, N_\downarrow)$-fermion sector and those in the ${(L-N_\uparrow, L-N_\downarrow)}$-fermion sector. The mapping between these two sectors can be done via a modified Shiba transformation $\hat{J}$, defined as
\begin{align}
\hat{J} &= \hat{J}_\uparrow \hat{J}_\downarrow, \\
\hat{J}_\sigma &= \prod_{j=1}^L \left[ \hat{c}^\dagger_{j,\sigma} - (-1)^j \hat{c}_{j,\sigma} \right] \hat{P}_\sigma.
\end{align}
Here $\hat{P}_\sigma$ is the lattice reflection operator that maps each lattice configuration to one in which fermions of spin $\sigma$ are moved from site $j$ to site $L-j+1$, and also multiplies by a factor of $(-1)$ for each pair of fermions swapped between sites.

Like the standard Shiba transformation in the homogeneous Hubbard model~\cite{essler2005}, this modified transformation maps fermions in low-energy states to fermionic holes in high-energy states; it includes the $(-1)^j$ phase factor and the spatial reflection which map between low- and high-energy orbitals. The transformation maps a state $|\Psi\rangle$ of $(N_\uparrow, N_\downarrow)$ fermions to a state $\hat{J} |\Psi\rangle$ of $(L-N_\uparrow, L-N_\downarrow)$ fermions. It is unitary and leaves the Hamiltonian invariant up to a constant: $\hat{J} \hat{H} \hat{J}^\dagger = \hat{H} + \mathrm{const}_0 + \mathrm{const}_1 \times \hat{N}$, which for a fixed $N$ reduces to a single constant shift. Consequently, if $|\Psi\rangle$ is the $N$-fermion ground state, then $\hat{J} |\Psi\rangle$ is the ground state of the ($2L-N$)-fermion sector, so the modified Shiba transformation provides direct access to over-half-filled ground states from their under-half-filled counterparts.

\section{Estimating $\beta_{c1}$ in the phenomenological model}
\label{sec:phenomenological-model-estimate-betac1}

Our phenomenological model (Sec.~\ref{sec:phenomenological-model}) predicts that pair breaking is governed by the successive equalization of single-particle orbital energies with bound-pair energies. In Sec.~\ref{sec:one-body-spectrum}, we provided a numerically found estimate [Eq.~\eqref{eq:lowest-energy-approximation}] for the lowest single-particle orbital energy. We now use this to estimate the threshold gradient $\beta_{c1}$ at which the first pair breaks into two fermions occupying single-particle orbitals.

Substituting Eq.~\eqref{eq:lowest-energy-approximation} into Eq.~\eqref{eq:numerical-estimate-for-betac1}, we obtain
\begin{equation}
\label{eq:equation-for-beta-c1}
    2 \left[ - \beta_{c1} \frac{L+1}{2} -2 + 2.270 \,\beta_{c1}^{2/3} \right] \approx -\sqrt{U^2+16}.
\end{equation}

To approximately solve Eq.~\eqref{eq:equation-for-beta-c1} for $\beta_{c1}$, we can rewrite it as a cubic equation for $y = \sqrt[3]{(L+1)\beta_{c1}}$,
\begin{equation}
\label{eq:cubic-equation-for-beta-c1}
    y^3 - \frac{2 \cdot 2.270}{(L+1)^{2/3}}\, y^2 - \left[ \sqrt{U^2+16} - 4 \right] = 0.
\end{equation}
In the large-$L$ limit, this has the real solution $y = \left( \sqrt{U^2+16} - 4 \right)^{1/3} \equiv y_0$. Expanding $y$ in powers of $(L+1)^{-2/3}$,
\begin{equation}
    \label{eq:y-expanded}
    y = y_0 + y_1\, (L+1)^{-2/3} + O\left(L^{-4/3}\right),
\end{equation}
and substituting into Eq.~\eqref{eq:cubic-equation-for-beta-c1}, we find $y_1 = \frac{2}{3} \cdot 2.270$.

Calculating $\beta_{c1}(L+1) = y^3$ from Eq.~\eqref{eq:y-expanded} then yields
\begin{align}
\beta_{c1} (L+1) &= \left(\sqrt{U^2+16} -4 \right) \\
&+ \left(\sqrt{U^2+16} -4\right)^{2/3} \frac{2\cdot 2.270}{(L+1)^{2/3}} + O\left(L^{-4/3}\right). \nonumber
\end{align}

\section{Grand-canonical phase diagram of the attractive Hubbard model}
\label{sec:phase-diagram}

\begin{figure}[t]
    \centering
    \includegraphics[width=0.93\linewidth]{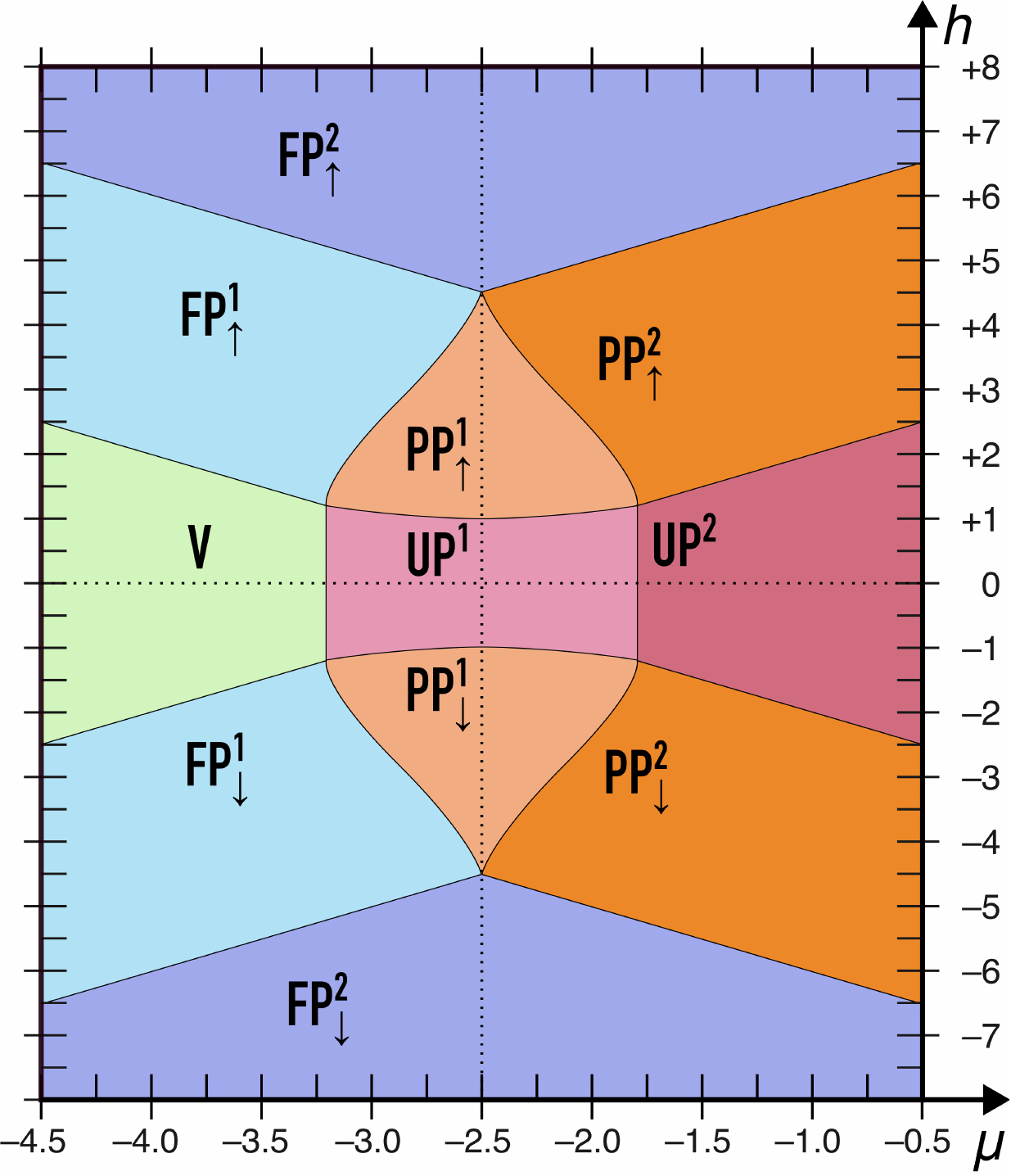}
    \caption{The grand-canonical phase diagram of the Hubbard model in the $\mu$--$h$ plane, for $U=-5$. Regions are labeled with phase names: $\mathrm{FP}$ = fully polarized, $\mathrm{PP}$ = partially polarized, $\mathrm{UP}$ = unpolarized, $\mathrm{V}$ = vacuum. Superscripts $1$ and $2$ denote, respectively, phases without and with a fully filled band. Arrow subscripts $\uparrow$, $\downarrow$ indicate which spin is the majority in that region. The depicted phase boundaries are derived from numerical calculations, as detailed in the text.}
    \label{fig:hubbard-grand-canonical-phase-diagram}
\end{figure}

Detailed discussions of the phase diagram of the attractive Fermi--Hubbard model can be found, for example, in Refs.~\cite{PhysRevA.81.053602,essler2005}. Here we provide only a brief overview and describe our method for numerically estimating the density at each point $(\mu, h)$.

A phase diagram of the attractive Hubbard model in the variables $\mu$ and $h$ is shown in Fig.~\ref{fig:hubbard-grand-canonical-phase-diagram} for $U = -5$. Several phases can be defined through the ground-state densities $n_\sigma(\mu,h;U)$: phase $\mathrm{UP}$---unpolarized (${n_\uparrow = n_\downarrow > 0}$); phase $\mathrm{PP}_\sigma$---partially polarized (${n_\sigma > n_{-\sigma} > 0}$); phase $\mathrm{FP}_\sigma$---fully polarized (${n_\sigma > 0, n_{-\sigma} = 0}$); and phase $\mathrm{V}$---vacuum (${n_\uparrow = n_\downarrow = 0}$). The superscripts $1$ and $2$ distinguish phases without filled bands from those in which at least one component has a fully filled band ($n_\sigma = 1$). The subscripts $\uparrow$ and $\downarrow$ indicate the majority spin ($\uparrow$ for $h>0$, $\downarrow$ for $h<0$).

For some regions of this phase diagram, the ground-state equations of state $n_\sigma(\mu,h;U)$ and $ n_D(\mu,h;U)$ can be given in closed form. However, within the $\mathrm{UP}$ and $\mathrm{PP}$ regions, these equations cannot be written as simple closed-form expressions~\cite{essler2005}. We therefore estimate the densities throughout the phase diagram numerically via DMRG calculations, as follows.

For a uniform Hubbard lattice, finite but with a large number of sites $L'$ (here set to $L'=160$), we perform DMRG calculations at different fixed fermion numbers ($L' \ge N_\uparrow \ge N_\downarrow \ge 0$;  $N_\uparrow+N_\downarrow\le L'$) to find the ground-state energies $E$ and doublon numbers $N_D$ corresponding to various densities $(n_\uparrow,n_\downarrow) = ( N_\uparrow/L', N_\downarrow/L')$. Then, the values $(\mu,h)$ corresponding to each pair $(n_\uparrow,n_\downarrow)$ are found via the following derivatives (estimated as finite differences):
\begin{align}
    \mu &= \left( \frac{\partial e}{\partial n} \right)_s, \\
    h &= \left( \frac{\partial e}{\partial s} \right)_n,
\end{align}
where ${e = E/L'}$, ${n = n_\uparrow + n_\downarrow}$, and ${s = n_\uparrow - n_\downarrow}$. This procedure yields a dense grid of data points $(\mu,h,n_\uparrow,n_\downarrow,e,n_D)$, with the doublon density estimated as ${n_D = N_D/L'}$.

To obtain data points corresponding to $N_\downarrow > N_\uparrow$ or $N_\uparrow+N_\downarrow > L'$, we exploit the spin and particle--hole symmetries of the Hubbard model, which give the following relations:
\begin{align}
n_\uparrow(\mu,-h) &= n_\downarrow(\mu,h), \\
n_\downarrow(\mu,-h) &= n_\uparrow(\mu,h), \\
n_\uparrow\left(\mu_0+\mu,h\right) &= 1 - n_\downarrow\left(\mu_0-\mu,h\right), \\
n_\downarrow\left(\mu_0+\mu,h\right) &= 1 - n_\uparrow\left(\mu_0-\mu,h\right),
\end{align}
where $\mu_0 = U/2$ is the chemical potential where the system is at half-filling. The doublon density $n_D = \langle \hat{n}_{j,\uparrow} \hat{n}_{j,\downarrow} \rangle$ transforms accordingly: 
\begin{align}
n_D(\mu,-h) &= n_D(\mu,h), \\
n_D(\mu_0+\mu,h) &= 1 - n(\mu_0-\mu,h) + n_D(\mu_0-\mu,h).
\end{align}

Note that this approach does not yield uniform coverage of the entire phase diagram. For example, inside the $\mathrm{UP}$ region, the densities $n_\uparrow = n_\downarrow$ depend only on $\mu$ and not on $h$. When using the DMRG approach described above, any pair of identical densities $(n_\uparrow = n_\downarrow)$ ends up assigned only to points $(\mu,h)$ that lie on the $\mathrm{UP}$--$\mathrm{PP}_\uparrow$ boundary, leaving the intermediate $h$ values undefined. However, densities for points inside the $\mathrm{UP}$ region can then be trivially extrapolated from the boundary.

In the main text, when computing the density in the finite lattice via the LDA, we evaluate the density at any given $(\mu,h)$ by linear interpolation between grid points. Because within each phase region the densities vary continuously with $\mu$ and $h$, the linear interpolation yields good results.

\section{Estimating $\beta_{c2}$ in the local-density approximation}
\label{sec:lda-betac2-estimate}

In the Hubbard model phase diagram (Fig.~\ref{fig:hubbard-grand-canonical-phase-diagram}), the boundary between the $\mathrm{UP}^1$ and $\mathrm{V}$ regions lies at $\mu_c = -\frac{1}{2} \sqrt{U^2 +16}$~\cite{essler2005}. If the system is in the spin-separated regime, its corresponding LDA trajectory lies entirely within the $\mathrm{V}$, $\mathrm{FP}_\sigma^1$, and $\mathrm{FP}_\sigma^2$ regions. Assuming that the system is spin-balanced (in which case the trajectory must cross $h=0$), this trajectory must lie at ${\mu \le \mu_c}$, and the threshold $\beta_{c2}$ can therefore be identified as the gradient for which ${\mu = \mu_c}$.

In the $U = 0$ limit, the phase boundaries and the densities $n_\sigma(\mu, h; U)$ in the relevant phase regions are given by closed-form expressions~\cite{essler2005}:
\begin{equation}
\nonumber
\begin{array}{c|c|c}
\mathrm{V}                   &  |h| < -\mu - 2             & n_\uparrow = n_\downarrow = 0 \\
\hline \mathrm{FP}^1_\sigma  &  -\mu - 2 < |h| < -\mu + 2  & \begin{array}{l} \;\; n_\sigma = \frac{1}{\pi} \arccos\!\left(-\frac{\mu+|h|}{2}\right), \\ n_{-\sigma} = 0 \end{array} \\
\hline \mathrm{FP}^2_\sigma  &  -\mu + 2 < |h|             & n_\sigma = 1, \quad n_{-\sigma} = 0.
\end{array}
\end{equation}
In these regions, at most one spin species is present, so the interaction term plays no role and the density susceptibility $\partial n_\sigma/\partial U$ is zero. Hence, away from the boundaries with partially-polarized regions, the above densities and phase boundaries remain unchanged for $U \ne 0$. We can therefore establish exact relationships between $U$, $L$, $N_\sigma$, $\beta$, and $\mu$ within the ${\mu \le \mu_c}$ region.

Consider a spin-balanced system with ${\mu \le \mu_c}$, for which the two endpoints of the LDA trajectory, $(\mu, h(1)) = (\mu, \beta[L-1]/2)$ and $(\mu, h(L)) = (\mu, -h(1))$, lie in opposite $\mathrm{FP}^2_\sigma$ regions. Then the total up-spin density $N_\uparrow$ along the trajectory can be estimated by replacing the sum over lattice sites with an integral over $h$:

\begin{align}
N_\uparrow &= \sum_j n_{\uparrow}(\mu, h(j);U) \approx \frac{1}{\beta} \int^{h(1)}_{0} \mathrm{d}h \; n_\uparrow(\mu, h;U) \nonumber \\
&= \frac{1}{\beta} \left[ \int_{-\mu-2}^{-\mu+2} \mathrm{d}h \; \frac{1}{\pi} \arccos\!\left(-\frac{\mu+|h|}{2}\right) + \int_{-\mu+2}^{h(1)} \mathrm{d}h \; 1 \right] \nonumber \\
&= \frac{2}{\beta} + \frac{h(1) + \mu - 2}{\beta} = \frac{L-1}{2} + \frac{\mu}{\beta}.
\end{align}

Given the LDA symmetries, we immediately have $N_\downarrow = N_\uparrow = N/2$. The overall relationship is
\begin{equation}
N \approx L-1 + \frac{2\mu}{\beta}.
\end{equation}

Inverting this relation yields
\begin{equation}
\beta \approx \frac{2 \mu}{N-L+1}.
\end{equation}

Setting $\mu = \mu_c = -\frac{1}{2}\sqrt{U^2 + 16}$, we obtain the relationship
\begin{equation}
\label{eq:betac2_lda_estimate_largepop}
\beta_{c2} \approx \frac{\sqrt{U^2+16}} {L-N-1}.
\end{equation}

The estimate in Eq.~\eqref{eq:betac2_lda_estimate_largepop} is valid provided the point $h(1)$ lies inside the $\mathrm{FP}^2_\uparrow$ region, \emph{i.e.}, $h(1) \ge -\mu + 2$, which means $N_\uparrow$ is larger than $\frac{1}{\beta}\int_{-\mu-2}^{-\mu+2} \mathrm{d}h \; \pi^{-1}\arccos\!\left(-\frac{\mu+|h|}{2}\right) = \frac{2}{\beta}$. For smaller populations, $h(1)$ lies inside the $\mathrm{FP}^1_\uparrow$ region. The total density $N_\uparrow$ then satisfies
\begin{equation}
N_\uparrow \approx \frac{1}{\beta}\int_{-\mu-2}^{h(1)} \mathrm{d}h \; \frac{1}{\pi} \arccos\!\left(-\frac{\mu+|h|}{2}\right).
\end{equation}

Evaluating this integral gives
\begin{align}
N_\uparrow \approx \frac{2}{\beta\pi}\left[ \lambda \arccos\!\left( -\lambda \right) + \sqrt{1-\lambda^2} \right] \equiv \frac{2}{\beta\pi} f(\lambda),
\end{align}
where we have defined $\lambda \equiv \frac{\beta (L-1)}{4} + \frac{\mu}{2} = \frac{h(1)}{2} + \frac{\mu}{2}$. In phase diagram terms, $2\lambda$ is the difference between $h(1)$ and the midpoint $h=-\mu$ of the $\mathrm{FP}^1_\uparrow$ region. The function $f(\lambda)$ is monotonically increasing in $\lambda$, and takes real values only in the ${-1 < \lambda < 1}$ interval (this is equivalent to the condition ${-\mu - 2 < h(1) < -\mu + 2}$).

To express $\lambda$ in terms of $f(\lambda)$, we expand $f(\lambda)$ in a Taylor series around $\lambda = 0$:
\begin{equation}
    f(\lambda) = \lambda \arccos\!\left( -\lambda \right) + \sqrt{1-\lambda^2}  = 1 + \frac{\pi \lambda}{2} + \frac{\lambda^2}{2} + O\left(\lambda^4\right).
\end{equation}

Truncating at quadratic order yields the following equation for $\lambda$:
\begin{equation}
    \lambda^2 + \lambda \pi + 2 [1 - f(\lambda)] = 0,
\end{equation}
with solution
\begin{equation}
    \lambda \approx -\frac{\pi}{2} \pm \frac{\sqrt{\pi^2-8+8 f(\lambda)}}{2},
\end{equation}
where the positive square root must be selected, so as to satisfy $-1 < \lambda < 1$.

Substituting $f(\lambda) = \frac{\beta \pi}{2} N_\uparrow $ and $\lambda = \frac{\beta (L-1)}{4} + \frac{\mu}{2}$, we obtain
\begin{equation}
\label{eq:relationship-between-beta-l-mu-nup}
    \mu + \pi + \frac{1}{2}\beta (L-1) \approx \sqrt{\pi^2 - 8 + 4\pi \beta N_\uparrow},
\end{equation}
which is the approximate relationship between $\beta$, $L$, $\mu$, and $N_\uparrow$.

Squaring both sides of Eq.~\eqref{eq:relationship-between-beta-l-mu-nup} yields a quadratic equation for $\beta$:
\begin{align}
    \beta^2 \cdot  &\frac{(L-1)^2}{4}   \\
    + \beta \cdot &\left[ (L-1)(\mu+\pi)  - 4 \pi N_\uparrow\right] \nonumber \\
    + &\left[ (\mu + \pi)^2 - \pi^2 + 8\right] = 0. \nonumber
\end{align}

Setting ${\mu = \mu_c = -\frac{1}{2}\sqrt{U^2 + 16}}$ gives the equation for $\beta_{c2}$:
\begin{align}
    \beta_{c2}^2 \cdot  &\frac{(L-1)^2}{4}   \\
    + \beta_{c2} \cdot &\left[ (L-1)\left(\pi-\frac{1}{2}\sqrt{U^2 + 16}\right)  - 4 \pi N_\uparrow\right] \nonumber \\
    + &\left[ 12 + \frac{U^2}{4} - \pi\sqrt{U^2+16} \right] = 0 \nonumber
\end{align}

The estimated $\beta_{c2}$ is then obtained as the appropriate root of this quadratic equation. Specifically, the root must satisfy ${ \frac{\sqrt{U^2+16}-4}{L-1} <  \beta_{c2} < \frac{\sqrt{U^2+16}+4}{L-1} }$ (due to the condition ${-\mu - 2 < h(1) < -\mu + 2}$), as well as the small-population condition $\beta_{c2} < 2/N_\uparrow$.

\section{Signatures of Fulde-Ferrell--Larkin-Ovchinnikov Cooper pairing in the pair correlations}
\label{sec:fflo-signatures}

Trapped Fermi gases are of interest to solid-state physicists in part because they can realize phases that mimic the Cooper pairing of solid-state superconductors~\cite{kinnunen2018fulde}. Quasi-one-dimensional systems are of particular interest, since they can stably support the exotic Fulde-Ferrell--Larkin-Ovchinnikov (FFLO) superconducting phase~\cite{PhysRevB.76.220508}. FFLO is a form of superconductivity which can occur when the two opposite-spin components that form the Cooper pairs have mismatched Fermi surfaces; this occurs, for example, for an imbalanced spin population~\cite{fulde1964superconductivity,1965-Larkin-Soviet}. Its defining feature is a nonzero center-of-mass momentum $q$ of the pairs, equal to the mismatch between the Fermi momenta, in contrast to standard Bardeen-Cooper-Schrieffer (BCS) superconductivity, where $q=0$.

In the attractive Hubbard chain, the FFLO-like state manifests itself as an oscillation in the pairing correlations with wave number $q$; on a finite, open chain of $L$ sites, we take $q = \pi |N_\uparrow-N_\downarrow|/(L+1)$, which is the finite-size form of the Fermi momentum mismatch~\cite{Rizzi2008}. Although the systems studied in this work are globally balanced, for $\beta_{c1} < \beta < \beta_{c2}$ their ground states contain extended regions of spatially varying local spin imbalance, which raises the question of whether FFLO-type correlations can appear locally, as they do in other inhomogeneous but balanced settings~\cite{PhysRevResearch.3.043105}. The local-density picture of Sec.~\ref{sec:lda} makes this expectation concrete: the partially polarized buffer regions map onto the PP phase of the homogeneous model [Fig.~\ref{fig:hubbard-grand-canonical-phase-diagram}], whose pair correlations are of the FFLO type~\cite{HeidrichMeisner2010}, with a local pair momentum set by the local spin imbalance. In this appendix, we test this expectation directly on the DMRG ground states.

In the strongly attractive regime, where bound pairs can be approximated as composite hardcore bosons, Cooper pairing manifests itself as a condensation of the bosonic fluid---\emph{i.e.} the macroscopic occupancy of a single eigenorbital of the doublon density matrix $\hat{\rho}^D$,
\begin{equation}
    \label{eq:doublon-rdm}
    \rho^D(i,j) = \left\langle \hat{c}^\dagger_{i,D} \hat{c}_{j,D} \right\rangle,
\end{equation}
where $\hat{c}_{j,D} = \hat{c}_{j,\downarrow} \hat{c}_{j,\uparrow}$ and, in finite systems, $\mathrm{Tr}\, \hat{\rho}^D = N_D$. In one dimension, true condensation---for which the pair correlations remain finite at arbitrarily long distances---is not possible, but one obtains instead a quasi-condensate, in which the correlations decay as a power law in $|i-j|$~\cite{PhysRevB.63.140511,PhysRevB.76.220508}. By analogy with a true condensate, in the quasi-condensed system the dominant (\emph{i.e.} largest-eigenvalue) orbital $\chi_1(j)$ of $\hat{\rho}^D$ can be taken as reflecting the underlying superfluid (pairing) order parameter~\cite{PhysRevB.76.220508,HeidrichMeisner2010}. FFLO pairing is then signalled by an oscillation of $\chi_1(j)$ with wave number $q$. In this appendix, we will use $\chi_1(j)$ as the primary diagnostic tool for occurrence of FFLO.

Before turning to the non-uniform system with $\beta > 0$, we establish a point of reference by examining two uniform systems with $\beta = 0$ and $U=-4$: one with balanced populations ($N=45+45$) and one with imbalanced populations ($N=45+32$), which host BCS- and FFLO-style correlations, respectively. These examples show how the two types of pair correlation can be distinguished through the orbital $\chi_1(j)$ of $\hat{\rho}^D$.

\begin{figure}
    \centering
    \includegraphics[width=0.95\linewidth]{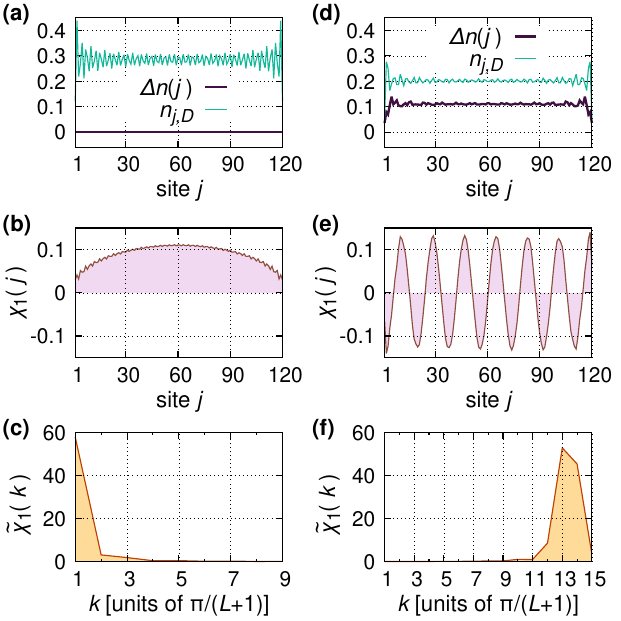}
    \caption{Example of Cooper pair correlations present in a uniform system ($\beta = 0$, $L=120$, $U=-4$), either spin-balanced [$N=45+45$, (a)-(c)] or spin-imbalanced [$N=45+32$, (d)-(f)]. (a),(d) The local spin imbalance $\Delta n(j) = |n_{j,\uparrow} - n_{j,\downarrow}|$ and the doublon density $n_{j,D}$. In (a), $\Delta n(j)$ is zero everywhere. (b),(e) The dominant orbital $\chi_1(j)$ of the doublon density matrix $\hat{\rho}^D$. (c),(f) The squared absolute Fourier transform of $\chi_1(j)$ [Eq.~\eqref{eq:fourier_transform_of_chi1}]. Lines are only to guide the eye.}
    \label{fig:uniform-systems-pairing-correlations-example}
\end{figure}

Figure~\ref{fig:uniform-systems-pairing-correlations-example} shows the ground state of both systems, with $N=45+45$ in the left column and $N=45+32$ in the right. The first row shows the local spin imbalance $\Delta n(j) = |n_{j,\uparrow} - n_{j,\downarrow}|$ and the doublon density $n_{j,D}$. For $N=45+45$, $\Delta n(j)$ vanishes identically everywhere. For the imbalanced system, $\Delta n(j)$ oscillates around a baseline of $(N_\uparrow-N_\downarrow)/L \approx 0.11$. Note that, in general, FFLO pairing is expected to produce $|N_\uparrow-N_\downarrow|$ maxima in $\Delta n(j)$, corresponding to the excess majority fermions gathering in the nodes of the pairing order parameter. For the case examined here, $\Delta n(j)$ does exhibit weak maxima that roughly correspond to node positions, but they are too indistinct to serve as an indicator of FFLO pairing.

The second row of Fig.~\ref{fig:uniform-systems-pairing-correlations-example} shows the orbital $\chi_1(j)$. For $N=45+45$ [Fig.~\ref{fig:uniform-systems-pairing-correlations-example}(b)], the sign of the orbital does not change, while its amplitude varies because it is shaped by the finite lattice and vanishes at the edges. The constant sign is the expected signature of BCS pairing, for which $q=0$ and the pairing order parameter does not oscillate. For $N=45+32$ [Fig.~\ref{fig:uniform-systems-pairing-correlations-example}(e)], by contrast, $\chi_1(j)$ oscillates with a single wave number, and can be approximately described as $\propto \cos(q j)$. The orbital has 13 equally spaced nodes, which identifies the wave number as the predicted $q = \pi |N_\uparrow-N_\downarrow|/(L+1)$. The orbital $\chi_1(j)$ thus directly demonstrates FFLO pairing with momentum $q$.

A common diagnostic for FFLO pairing correlations is analyzing the pair momentum distribution, \emph{i.e.} the diagonal of the two-dimensional Fourier transform of $\rho^D(i,j)$. A symptom of FFLO pairing is a maximum in the pair momentum distribution at a nonzero pair momentum $q$. Here, we approximate the quasi-condensate part of the pair momentum distribution by the squared absolute Fourier transform of the dominant orbital $\chi_1(j)$:
\begin{equation}
\label{eq:fourier_transform_of_chi1}
\tilde{\chi}_1(k) = \left|\sum_{j = 1}^{L} \exp(-ikj) \chi_1(j) \right|^2.
\end{equation}
We perform this Fourier transform in the discrete, finite-lattice basis of quasimomenta, $k_m = m\pi/(L+1)$ with $m=1,2,\ldots,L$. For the two shown reference systems, we show the function $\tilde{\chi}_1(k)$ in the third row of Fig.~\ref{fig:uniform-systems-pairing-correlations-example}. In both cases, $\tilde{\chi}_1(k)$ serves as an immediate diagnostic to estimate the pairing momentum $q$; however, due to the finite-size boundary effects, its maximum does not necessarily fall precisely at $k_m = q$. For the $N=45+45$ case [Fig.~\ref{fig:uniform-systems-pairing-correlations-example}(c)], the Fourier transform has a maximum at $k = 1 \times \pi/(L+1)$ (the smallest possible $k_m$), while for $N=45+32$ [Fig.~\ref{fig:uniform-systems-pairing-correlations-example}(f)] the Fourier transform has a maximum that straddles ${k = 13 \pi/(L+1)}$ and ${k = 14\pi/(L+1)}$.

With these reference cases established, we now turn to a system with $\beta > 0$, in order to see how the orbital $\chi_1(j)$ changes and whether it displays any FFLO-like features.

\begin{figure}
    \centering
    \includegraphics[width=0.95\linewidth]{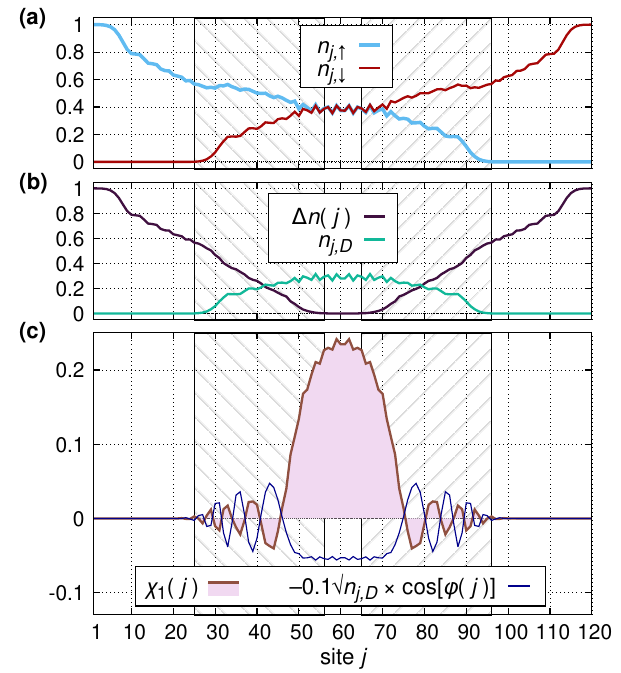}
    \caption{Density distribution and pairing order parameter in a system with parameters: $L = 120, N_\uparrow = 45, N_\downarrow = 45, U = -4, \beta = 0.08$. The gray areas denote the site ranges $25 \le j \le 56$, $65 \le j \le 96$, for which $\Delta n(j) \ge 10^{-3}$ and $\min[n_{j,\uparrow},n_{j,\downarrow}] \ge 10^{-3}$, which we interpret as the ``partially-polarized buffer region''. (a) The up- and down-spin density profiles $n_{j,\uparrow}$, $n_{j,\downarrow}$. (b) The spin imbalance $\Delta n(j) = |n_{j,\uparrow} - n_{j,\downarrow}|$ and doublon density $n_{j,D}$. (c) Thick line and filled curve area: the dominant orbital $\chi_1(j)$ of the doublon density matrix $\hat{\rho}^D$. Thin line: the approximate prediction for FFLO-type pairing correlations, $\cos\left[\varphi(j)\right]$, where $\varphi(j)$ [Eq.~\eqref{eq:total_accumulated_fflo_phase}] is the phase accumulated relative to the middle of the lattice. The latter function is multiplied by an envelope $-0.1 \sqrt{n_{j,D}}$, so that the locations of its oscillation nodes can be more easily compared with $\chi_1(j)$.}
    \label{fig:example-of-non-uniform-system-with-fflo}
\end{figure}

Figure~\ref{fig:example-of-non-uniform-system-with-fflo} shows a system with $N=45+45$, $U=-4$, and $\beta=0.08$. Panels (a) and (b) show the densities $n_{j,\sigma}$, the local spin imbalance $\Delta n(j)$, and the doublon density $n_{j,D}$. The system is phase-separated into an unpolarized core, polarized wings, and a partially polarized transition region (buffer) straddling the two. We define the buffer region as the region in which both the spin imbalance and the minority spin density are non-negligible, $\ge 10^{-3}$, which here gives the site ranges ${25 \le j \le 56}$ and ${65 \le j \le 96}$; these ranges are marked by the shaded areas. (Since a single fermion contributes an average density of ${L^{-1} \sim 10^{-2}}$ per site, a threshold one order of magnitude lower is a reasonable criterion for negligible density.) Throughout the buffer region, $\Delta n(j)$ increases continuously from the core towards the wings.

Panel (c) shows the dominant orbital $\chi_1(j)$ (thick line, filled shape). We find that the orbital displays different features within different regions of the lattice. Within the core region ($57 \le j \le 64$), the sign of $\chi_1(j)$ is constant, similarly as it was in Fig.~\ref{fig:uniform-systems-pairing-correlations-example}(b) for the balanced system, although the amplitude follows an envelope set by the doublon density; this indicates that pairing in the unpolarized core is of the zero-momentum BCS type. As we move further into the buffer region, $\chi_1(j)$ begins to oscillate, signalling locally present FFLO pairing. As we move towards the wings and leave the buffer region, $\chi_1(j)$ decays rapidly to zero. Note that, similarly to Fig.~\ref{fig:uniform-systems-pairing-correlations-example}(d), the FFLO-like features cannot be easily detected via oscillations in the one-body $\Delta n(j)$ observable. While $\Delta n(j)$ does show some oscillations in the FFLO-like region, these maxima are very indistinct.

These oscillations suggest FFLO pairing, but it remains to be checked whether their wave numbers matches that of the theoretically expected order parameter. No single FFLO momentum $q$ can be assigned within the buffer region, because $\Delta n(j)$ varies monotonically rather than oscillating around a fixed baseline. Instead, a local-density-approximation argument suggests that the pair momentum corresponding to each site follows the local spin imbalance, and we estimate it as $q(j) = \pi \Delta n(j) L/(L+1)$; that is, the local pair momentum equals the value it would take in a lattice with total population imbalance $L \Delta n(j)$. The pairing order parameter is then expected to vary in space as $\cos[\varphi(j)]$, where the local phase $\varphi(j)$ is an integral over $q(j)$. To find $\varphi(j)$, we treat the middle of the lattice (the point corresponding to site number $(L+1)/2 = 60.5$) as the point of zero phase, and integrate the local $q(j')$ out to site $j$ using the trapezoidal rule. Explicitly, for $j > (L+1)/2$,
\begin{align}
\label{eq:total_accumulated_fflo_phase}
\varphi(L/2 + 1) &= \frac{1}{2} q(L/2+1), \nonumber \\
\varphi(j+1) &= \varphi(j) + \frac{q(j+1) + q(j)}{2},
\end{align}
and we use an analogous expression for $j < (L+1)/2$.

Figure~\ref{fig:example-of-non-uniform-system-with-fflo}(c) shows $\cos[\varphi(j)]$ (thin line), multiplied by an envelope $-0.1 \times \sqrt{n_{j,D}}$, for comparison with $\chi_1(j)$. Note that the envelope is not intended to reproduce the shape of $\chi_1(j)$; it merely mimics the decay of the pairing towards the lattice edges to make the visual comparison easier. The oscillation nodes of $\cos[\varphi(j)]$ agree very well with those of $\chi_1(j)$, indicating that the structure of $\chi_1(j)$ does follow the local spin imbalance, as expected for FFLO correlations.

For completeness, we also check whether the observed FFLO correlations leave a clear signal in the Fourier transform of $\chi_1(j)$, as was the case for the uniform systems depicted earlier [Fig.~\ref{fig:uniform-systems-pairing-correlations-example}(c,f)]. In order to isolate the pairing correlations of the buffer region, we define a windowed Fourier transform that restricts the sum to a chosen part of the lattice,
\begin{equation}
\label{eq:fourier_transform_of_chi1_windowed}
\tilde{\chi}_1(j_\mathrm{max} ; k) = \left|\sum_{j = 1}^{j_\mathrm{max}} \exp(-ikj) \chi_1(j) \right|^2.
\end{equation}

\begin{figure}
    \centering
    \includegraphics[width=0.95\linewidth]{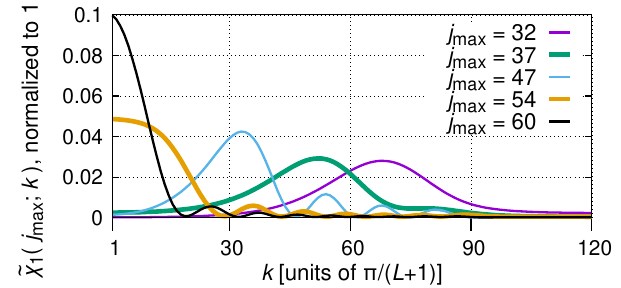}
    \caption{The Fourier transform [Eq.~\eqref{eq:fourier_transform_of_chi1_windowed}] calculated for the orbital $\chi_1(j)$ from Fig.~\ref{fig:example-of-non-uniform-system-with-fflo}(c), for increasingly larger windows $j = 1 \ldots j_\mathrm{max}$. Each curve has been normalized to unity.}
    \label{fig:windowed-fourier-transform-of-chi1}
\end{figure}

Figure~\ref{fig:windowed-fourier-transform-of-chi1} shows $\tilde{\chi}_1(j_\mathrm{max} ; k)$ for the orbital $\chi_1(j)$ of Fig.~\ref{fig:example-of-non-uniform-system-with-fflo}(c), evaluated for a series of increasingly large windows $j_\mathrm{max}$ that cover progressively more of the left buffer region. Given the oscillatory nature of $\chi_1(j)$, the maximum occurs at a finite center-of-mass momentum, as expected. Its position, however, depends on $j_\mathrm{max}$: each time the window is extended towards the lattice center, the newly included parts of $\chi_1(j)$ oscillate more slowly but have a larger magnitude, and so come to dominate the transform. Once the window reaches the core ($j_\mathrm{max} = L/2$), the BCS pairing of the core dominates and the maximum sits at $k = 1\pi/(L+1)$. Overall, we find that there is no particular dominant value of $q$ that is independent of the Fourier transform window.

In summary, this example shows that the $\beta > 0$ system can display evidence of FFLO-type correlations. Notably, the FFLO-like features appear even though the spin populations are globally balanced. However, these features are confined to the transitional partially-polarized region. Moreover, rather than a single well-defined $q$, the system hosts a broad distribution of local momenta $q(j)$.

\bibliography{_Biblio}

\end{document}